\documentclass[%
 reprint,
 amsmath,amssymb,
 aps,
 prl
]{revtex4-1}

\usepackage{xcolor}
\usepackage{graphicx}
\usepackage{dcolumn}
\usepackage{bm}
\usepackage{stmaryrd}
\usepackage{textgreek}

\begin{document}

\preprint{APS/PRL-correlation-0.1}

\title{A quantitatively consistent, scale-spanning model for same-material tribocharging}

\author{Galien Grosjean}%
 \email{galienmariep.grosjean@ist.ac.at}
\author{Sebastian Wald}
\author{Juan Carlos Sobarzo}
\author{Scott Waitukaitis}
\affiliation{%
 IST Austria\\
 Lab Building West\\
 Am Campus 1\\
 3400 Klosterneuburg AT
}%

\date{\today}

\begin{abstract}

We propose a quantitative, scale-spanning model for same-material tribocharging.  Our key insight is to account for mesoscale spatial correlations in donor/acceptor surface properties, which dramatically affect the macroscopic charge transfer and quantitatively reconcile previous inconsistencies related to the microscale.  We furthermore identify a viable mechanism by which the mesoscale features emerge, which may help constrain the list of donor/acceptor candidates.  As the only free-parameters in our model involve the atomic scale, data analyzed in light of it could help resolve the detailed mechanism of tribocharging.

\end{abstract}

\maketitle

Tribocharging, \textit{i.e.}~charge transfer between materials during contact \cite{Lacks:2019dx}, plays a critical role in natural phenomena \cite{Wurm:2019, Schrader:2018eh, Steinpilz:2020iy, Desch:2000, Berdeklis:2001ta}, industrial processes \cite{Baytekin:2013dq, Abbasi:2007cv}, and energy harvesting devices \cite{Kanik:2014bn, Musa:2018uy, Wang:2013bq}, yet resists interpretation.  One fundamental roadblock has been the inability to identify the atomic-scale mechanism, and in particular the charge carriers, \textit{i.e.}~ions \textit{vs.}~electrons \cite{Lacks:2019dx}.  An equally important roadblock is the lack of quantitative agreement (or even comparison) between experiments and theory.  Experimentally, issues such as the difficulty of measuring contact areas render much data qualitative (\textit{e.g.}~the sign or scale of charging)~\cite{Hu:2012jf,Waitukaitis:2013fwa,Waitukaitis:2014cg,Collins:2018,Lee:2018gg,Harris:2019}.  Theoretical advances are stymied by the multi-scale nature of the effect, where one must simultaneously account for probabilistic effects at the atomic scale ($<$1 nm), unexplained emergent features at the mesoscale ($\sim$1 \textmu m), and then through these explain the familiar behavior of the macroscale ($>$1 mm).  These challenges have perhaps led to a general disinterest in quantitative reconciliation, with some authors characterizing the outlook on achieving a scale-spanning description as `impossible' \cite{Lacks:2011hm}.

\begin{figure}[ht!]
\includegraphics[width=8.6cm]{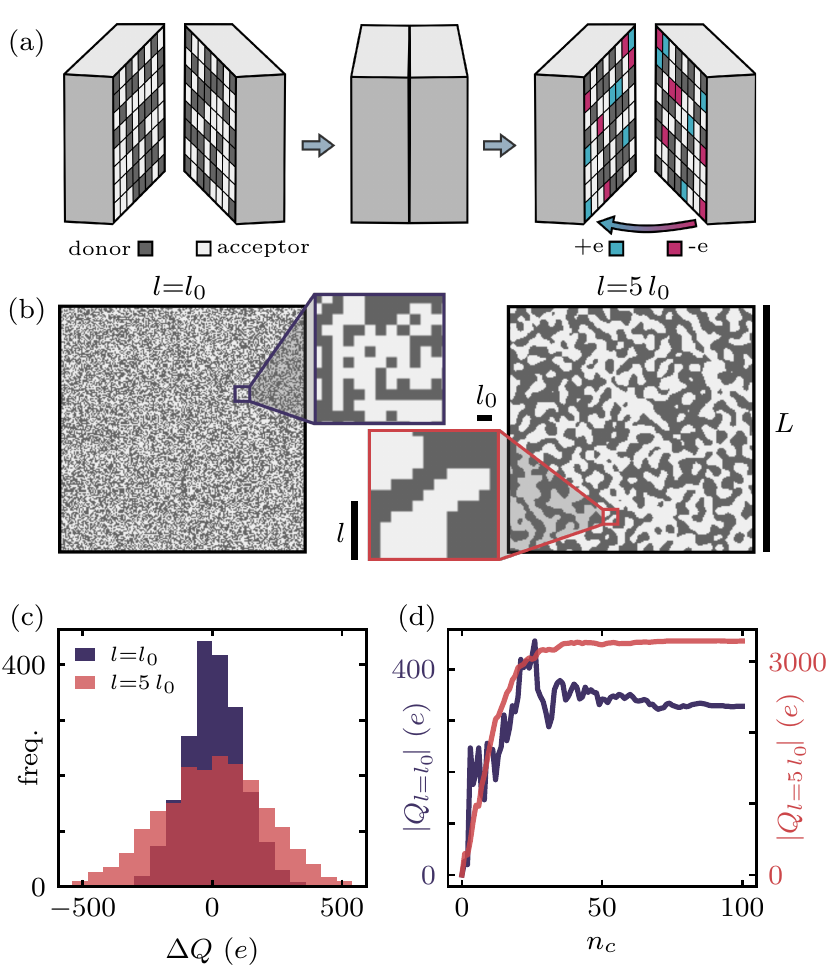}
\caption{\label{fig1} (a) Apodaca \emph{et al.} suggested that same-material tribocharging may arise from equally-sized, randomly-distributed donor/acceptor sites, where unit charges, $e$, are transferred during contact~\cite{Apodaca:2009dr}.  (b)  Surfaces with identical microscopic size $L$, microscopic scale $l_0$, but different mesoscale correlation lengths, $l$. On the left, $l\!=\!l_0$, and on the right $l\!=\!5\;l_0$ (where $l_0$ is the site size).  (c)  The single-contact distribution in $\Delta Q$ is broader when $l\!=\!5\;l_0$ (violet, $\sim$200 $e$) than when $l\!=\!l_0$ (pink, $\sim$100 $e$).  (d)  For sequential contacts, charge transfer is enhanced when $l\!=\!5\;l_0$, and fluctuations that dominate the $l\!=\!l_0$ case are suppressed.}
\end{figure}

Same-material tribocharging, where charge is exchanged between identical materials, is perhaps the most puzzling manifestation of the phenomenon.  It has been attributed to trapped electrons \cite{Lowell:2000jk,Lowell:2000bl, Lacks:2007ge, Lacks:2008hz, Duff:2008, Forward:2009in, Lacks:2016}, induced polarization \cite{Shinbrot:2017, Siu:2014, Kolehmainen:2018, Yoshimatsu:2016, Yoshimatsu:2016b}, or mechanochemistry \cite{Sow:2012, Sow:2012b, Sow:2013, Baytekin:2013dq}.  Yet, experiments on same-material tribocharging have produced some valuable clues.  Using soft (Young's modulus $\sim$1 MPa), atomically smooth (roughness $<1$ nm) polymers to achieve conformal contact, Apodaca \textit{et al.}~found that the magnitude of charge transfer grows with the square root of the contacting area \cite{Apodaca:2009dr}.  They proposed that the surface consists of equally-sized, randomly-assigned donor/acceptor sites (Fig.~\ref{fig1}a), each capable of giving/receiving one unit charge.  This allowed them to recover $\overline{|\Delta Q|}\!=\!C\sqrt{A}$, where the prefactor, $C$ depends on the length scale of a single site, $l_0$. Though $l_0$ should be on the scale of one atom, fits to their model required a value of $\sim$0.005~\r{A}---more than \textit{100 times smaller} than the Bohr radius of hydrogen.  Thus, while their idea is compelling, this discrepancy suggested it is still somehow a `toy model.'

Nonetheless, features suggestive of their idea have been observed, but at significantly larger scales.  Using Kelvin Force Probe Microscopy on the same polymers, Baytekin \textit{et al.}~found `mosaics'---regions of positive/negative charge after contact---that were spatially correlated over length scales up to $\sim$450 nm \cite{Baytekin:2011bx}.  This observation raises fundamental questions:  How can the impossibly small scales implied by Apodaca be connected to the mesoscopic ones seen by Baytekin?  What would the introduction of an intermediate scale change in Apodaca's model?  And why do the intermediate scale correlations emerge in the first place?  In this work, we use analysis supported by numerical simulations to investigate the multi-scale aspects of same-material tribocharging.  Our key development is to properly account for the mesoscale spatial correlations, which allows us to quantitatively resolve the discrepancies in the Apodaca framework and produce a scale-spanning model.

We start by explaining a first set of numerical simulations, where we mimic charge transfer between `synthetic' surfaces (in contrast to physically derived surfaces later) by creating two $N$-element matrices involving three length scales:  $l_0$, $l$, and $L$.  The smallest, $l_0$, corresponds to the elementary donors/acceptors of the atomic scale, and is represented by a single matrix element.  The largest, $L$, corresponds to the macroscopic system size.  We assume that there is a single intermediate scale, $l$, that characterizes the mesoscopic correlations observed by Baytekin.  (Note they measured charge, but this implies donor/acceptor correlations.)  Each matrix element is assigned as donor or acceptor, with probabilities $p$ and $1-p$, respectively. We account for correlations in assignments via thresholding a random scalar field (see Supplemental Material \cite{SupplMat}).  To perform a `contact', we first generate a `left' and `right' surface from identical input length scales and probabilities (Fig.~\ref{fig1}a).  Charge transfer of one unit, $e$, between matrix elements [\textit{i,j}] occurs if (1) [\textit{i,j}] on the left/right is a donor, (2) [\textit{i,j}] on the right/left is an acceptor, (3) the value of an independent random uniform variable is less than the transfer probability, $\alpha$, and (4) for sequential contacts, transfer at [\textit{i,j}] hasn't yet occurred.  The net charge transferred is the difference between left-to-right (`right') and right-to-left (`left') transfers.

Figure \ref{fig1}b shows two representative surfaces, one with $l\!=\!l_0$ and the second $l\!=\!5\;l_0$.  Although their only difference involves the length scale, $l$, we see stark changes in the charging behavior.  In Fig.~\ref{fig1}c, we plot distributions of the charge transferred in the first contact, $\Delta Q$, for 1000 pair-instances.  As we do not assume any post-contact discharge \cite{Haberle:2019}, both distributions are Gaussian and centered at zero, but while $l\!=\!l_0$ produces a width of $\sim$100 $e$, the $l\!=\!5\;l_0$ width is $\sim$200 $e$.  Correlations also change the behavior during sequential contacts.  Fig.~\ref{fig1}d shows two examples of the accumulated charge, $|Q|$, \textit{vs.}~the number of contacts, $n_c$.  Like the initial transfer, the final charge, $Q_f$, is typically larger when $l\!=\!5\;l_0$. Additionally, the fluctuations are on the order of $Q_f$ when $l\!=\!l_0$, but are hardly discernible when the $l\!=\!5\;l_0$ (consistent with experiments \cite{Apodaca:2009dr}).

\begin{figure}[t]
\includegraphics[width=8.6cm]{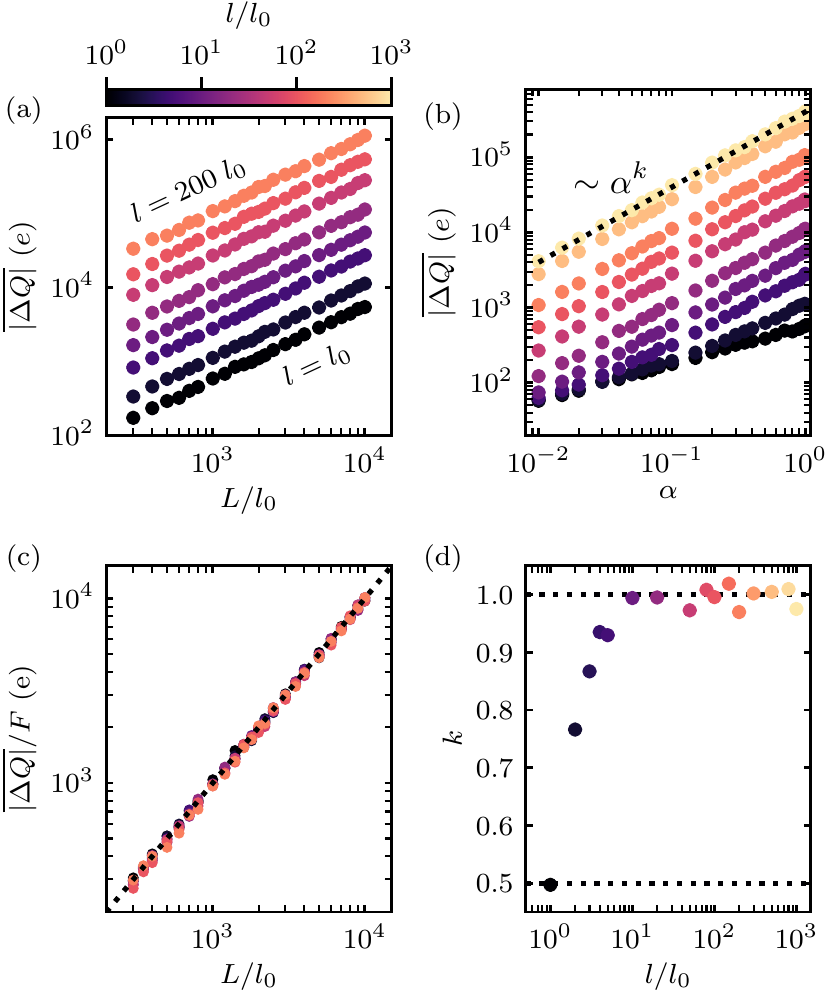}
\caption{\label{fig2} (a) Average charge transfer $\overline{|\Delta Q|}$ after one contact, for different values of $l$ and $L$ ($\alpha\!=\!1$, $p\!=\!0.5$). The scaling $\overline{|\Delta Q|}\!\propto\!L$ is recovered in all cases, but with a growing prefactor.  (b) For fixed $L$ and $l$, $\overline{|\Delta Q|}\!\propto\!\alpha^{k}$, where $k$ changes with $l$.  (c) The data from (a) for $l\!\gg\!l_0$ collapses to a line of slope unity when rescaled using Eq.~\ref{eq:mosaicScale}.  (d) The exponent $k$~\textit{vs.}~$l/l_0$.  When $l\! \approx\!l_0$, transfer is probabilistic (Eq.~\ref{eq:smallestScale}) and $\overline{|\Delta Q|}\!\propto\!\sqrt{\alpha}$.  When $l\!\gg\!l_0$, transfer occurs at a fixed rate (Eq.~\ref{eq:mosaicScale}) where $\overline{|\Delta Q|}\!\propto\!\alpha$.}
\end{figure}

We now examine the first contact behavior of these simulations in detail.  Figure \ref{fig2}a shows the average absolute value of charge exchanged, $\overline{|\Delta Q|}$, for increasing system size and several values of $l$.  The scaling $\overline{|\Delta Q|}\!\propto\!\sqrt{A}\!\propto\!L/l_0$ is recovered in all cases, but the prefactor steadily increases with $l\!>\!l_0$.  Figure~\ref{fig2}b shows that the dependence on the transfer probability, $\alpha$, exhibits unexpected non-linear behavior as a function of the correlation length, $l$. At every value of $l$, we see a trend consistent with a power law, \textit{i.e.}~$\overline{|\Delta Q|}\!\propto\!\alpha^{k}$.  However, the exponent, $k$, increases with $l$, starting at $k\!=\!0.5$ for $l\!=\!l_0$ and saturating at $k\!=\!1.0$ for $l\!\gg\!l_0$ (Fig.~\ref{fig2}d).  

To analytically explain these observations, we first consider the case $L\!\gg\!l=\!l_0$, here sketching our argument (details are in the Supplemental Material \cite{SupplMat}).  We momentarily focus on right transfers, which occur with compound probability $p(1-p)\alpha$.  Absent correlations, all sites [$i,j$] are independent, hence the total right charge transfer is Gaussian with mean $eNp(1-p)\alpha$ and width $e\sqrt{Np(1-p)\alpha(1-p(1-p)\alpha)}$.  A similar distribution exists for left transfer, but technically only when considered independently---simultaneous left/right transfer cannot occur.  Nonetheless, the probability for this is small, and we therefore approximate the left/right distributions as independent.  The net transfer, $\Delta Q$, is thus also Gaussian distributed, with zero mean and width $\sigma\!= \!e\sqrt{2Np(1-p)\alpha(1-p(1-p)\alpha)}$.  Neglecting terms like $\big{(}p(1-p)\alpha\big{)}^2$ and considering $\overline{|\Delta Q|}\!=\!\sqrt{2/\pi}\sigma$ \cite{SupplMat}, we find
\begin{equation}
    \overline{|\Delta Q|} = \sqrt{\frac{2}{\pi}}\frac{eL}{l_0}\sqrt{2p(1-p)\alpha}.
\label{eq:smallestScale}
\end{equation}
This recovers the $\sqrt{A}$ scaling, but points to a slight mistake in the earlier work \cite{Apodaca:2009dr} in that the $\alpha$-dependence is square root rather than linear---exactly as our simulations in Fig.~\ref{fig2}d.  In the Supplemental Material \cite{SupplMat}, we verify that Eq.~\ref{eq:smallestScale} collapses our simulated data for wide ranges of $p$ and $\alpha$.

Next we consider the case $L\!\gg\!l\!\gg\!l_0$ (again with details in the Supplemental Material \cite{SupplMat}).  This fundamentally alters the argument above as the site identities exhibit spatial correlations.  To handle this, we first imagine rescaling the system by $l/l_0$, leading to surfaces with $N'\!=\!N/(l/l_0)^2$ larger `patches', each consisting of many sites.  Identities of entire patches still occur with probabilities $p$ and $1-p$.  Next, we rescale back to deal with transfers, which still occur independently for each site.  During contact, regions of donors face acceptors with the characteristic size of a patch.  If the number of sites in these regions ($n\!=\!(l/l_0)^2$) is large, the mean transfer per patch ($\alpha n$) effectively hides the fluctuations ($\sqrt{n\alpha(1-\alpha)}$)---hence we treat $\alpha$ as a rate.  Thus, when $l\!\gg\!l_0$ we have
\begin{equation}
    \overline{|\Delta Q|} = \sqrt{\frac{2}{\pi}}\frac{e\alpha Ll}{l_0^2}\sqrt{2p(1-p)}.
\label{eq:mosaicScale}
\end{equation}
Here, like in the Apodaca work, the $\alpha$-dependence is linear.  The critical difference, however, is the dependence on the intermediate length scale, which \textit{amplifies} the charge transfer by the factor, $l/l_0$.  We confirm this with our simulated data in Fig.~\ref{fig2}c, where the prefactor $F\!=\!\alpha l/l_0 \sqrt{4p(1-p)/\pi}$ collapses $\overline{|\Delta Q|}$ when $l\!\gg\!l_0$.  Qualitatively, the explanation for this amplification is that the variability (\textit{i.e.}~standard deviation) in the number of donors/acceptors on a surface increases with the scale of spatial correlations.  One can quickly grasp why by considering the extreme case $l\!=\!L$, where each surface is either purely donor or acceptor, and consequently $\overline{|\Delta Q|}\!\propto\!\alpha e N\!\propto\!A$.  This highlights that same- and different-material tribocharging are two manifestations of a similar underlying phenomenon, only appearing different depending on the scale at which one looks. 

\begin{figure}[t]
\includegraphics[width=8.6cm]{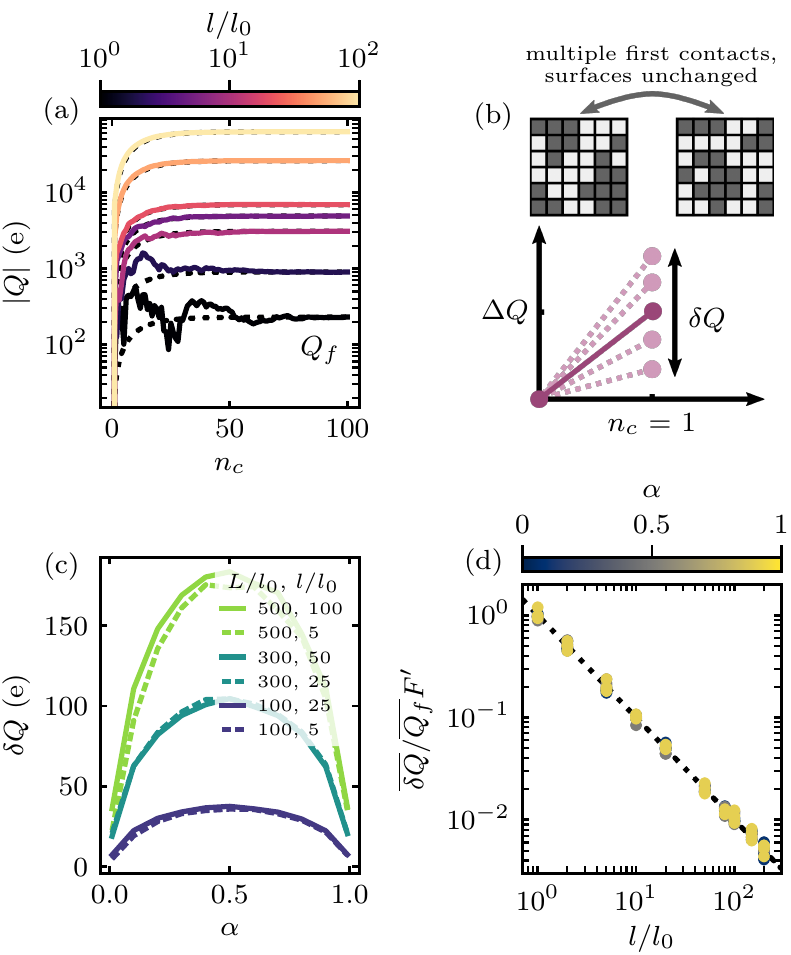}
\caption{\label{fig3} (a) Sequential contacts lead to accumulated charge, $|Q|$, which ultimately saturates at $Q_f$ (here for several $l/l_0$,  $\alpha\!=\!0.1$ and $p\!=\!0.5$).  For small $l/l_0$, fluctuations are on the order of $Q_f$, whereas for large $l$ they are suppressed.  Dotted lines correspond to Eq.~\ref{eq:exp} using the measured $Q_f$.  (b)  We quantify fluctuations by repeating the first contact between the same two surfaces and measuring the spread, $\delta Q$.  (c) For a particular surface pair, $\delta Q$ depends on both $\alpha$ and $L$, but is largely independent of $l$.  (d) Ensemble averages of relative fluctuations $\overline{\delta Q}/\overline{Q_f}$ collapsed by factor $F'\!=\! \sqrt{\alpha(1-\alpha)\pi/2}$ \textit{vs.}~$l/l_0$, which shows fluctuations are suppressed by spatial correlations.  Each point corresponds to the ensemble average over 20 pairs of surfaces of the fluctuations over 50 contacts.}
\end{figure}

We now turn to sequential contacts.  As the simulation results of Fig.~\ref{fig3}a show, repeated contacts with the same surfaces leads to curves in accumulated charge, $|Q|$ \textit{vs.}~$n_c$, that level off at some value, $Q_f$. In the Supplemental Material, we show that the underlying trend is given by a saturated exponential, \textit{i.e.},
\begin{equation}
    Q(n_c) = Q_f \left( 1 - \exp \left(-\alpha n_c \right) \right).
    \label{eq:exp}
\end{equation}
Figure \ref{fig3}a also illustrates the presence of fluctuations on approach to $Q_f$.  For large $l/l_0$, these aren't noticeable, but for small $l/l_0$ they are overwhelming, and cannot be suppressed even when we increase the system size, $L$.  We quantify their scale by repeatedly performing first contacts between individual surface pairs (`resetting' each time) and measuring the standard deviation, $\delta Q$ (Fig.~\ref{fig3}b).  In Fig.~\ref{fig3}c we show $\delta Q$ for a few pairs, which reveals that the fluctuations grow with the macroscale length, $L$, and depend strongly on $\alpha$, but are independent of the mesoscale, $l$.  To analyze why, we note that a particular pair has a fixed donor/acceptor arrangement, which means $\delta Q$ arises solely from $\alpha$.  Denoting the number of donors on the left/right that face acceptors on the right/left as $N_{\rightleftarrows}$, one finds that $\delta Q \!=\! e\sqrt{(N_\shortleftarrow + N_\shortrightarrow)\alpha(1-\alpha)}$.  In the Supplemental Material \cite{SupplMat}, we justify using the averages $\overline{N_\rightleftarrows}\!=\!(L/l_0)^2p(1-p)$ to find the ensemble expression, 
\begin{equation}
\overline{\delta Q} = e\frac{L}{l_0}\sqrt{2p(1-p)\alpha(1-\alpha)}.
\label{eq:first_flucts}
\end{equation}
This establishes that the fluctuations, like $\overline{|\Delta Q|}$, grow linearly with $L$, but unlike $\overline{|\Delta Q|}$ are independent of $l$.  Consequently, they cannot be suppressed by increasing system size, but can be suppressed by the introduction of the intermediate scale, $l$.  In Fig.~\ref{fig3}d, we collapse the simulated $\overline{\delta Q}$ data to our predicted line with the appropriate rescaling. 

The last question we posed remains:  Why do the intermediate scale correlations emerge in the first place?  A few mechanisms have been proposed. For example, with inelastically deformed materials it has been shown that micron-sized voids form, which are somewhat suggestive of mosaics~\cite{Wang:2017}.  Other authors make connections to `islands' of adsorbed surface water, but the sizes of these has not been measured nor compared to the charge mosaics~\cite{Xie:2016gr,  Yu:2017, Lee:2018gg, Harris:2019, Haberle:2019}.  In the case of void formation, the process for a growing lenghtscale is clear---the materials are ripped apart---yet the polymers used by Apodaca/Baytekin are highly elastic and not intentionally stretched.  In the case of water islands, and indeed virtually all `patch'-type models, one must assume features of a certain size exist, but how that length scale emerges is still unclear.

We propose that a nucleation process which energetically favors neighboring donors (or equivalently, neighboring acceptors) is a viable candidate.  We support this by developing a second, distinct set of simulations to mimic the physics of surface formation, which are time-dependent and latticed-based, and which again produce $L/l_0 \times L/l_0$ donor/acceptor matrices.  At each time step, we assume a donor site can transition into an acceptor site, and vice-versa. This could represent widely suspected mechanisms such as adsorption of an ambient donor species (\textit{i.e.}~H$_2$O), but also novel ideas such as phase separation during polymer curing.  The transition probabilities of a given site depend on its neighbors \textit{i.e.},
\begin{equation}
\begin{aligned}
    P_A (\nu) &= P_0 \exp \left( - K \nu \right)\\
    P_D (\nu) &= P_0 \exp \left( - K (4-\nu) \right),
\end{aligned}
\label{eq:proba}
\end{equation}
where $\nu$ is the number of neighbors that are donors (\textit{i.e.}~$\nu \in [0,4]$). The exponential form is motivated from an Arrhenius-like process where each neighbor modifies a local energy barrier by $\epsilon/kT\!=\!K$. Such as process could equally represent interactions between individual atoms or in a continuous medium~\cite{Brune:1998,Varanasi:2009,Lu:2018}.  Full details can be found in the Supplemental Material~\cite{SupplMat}. 

\begin{figure}[t]
\includegraphics[width=8.6cm]{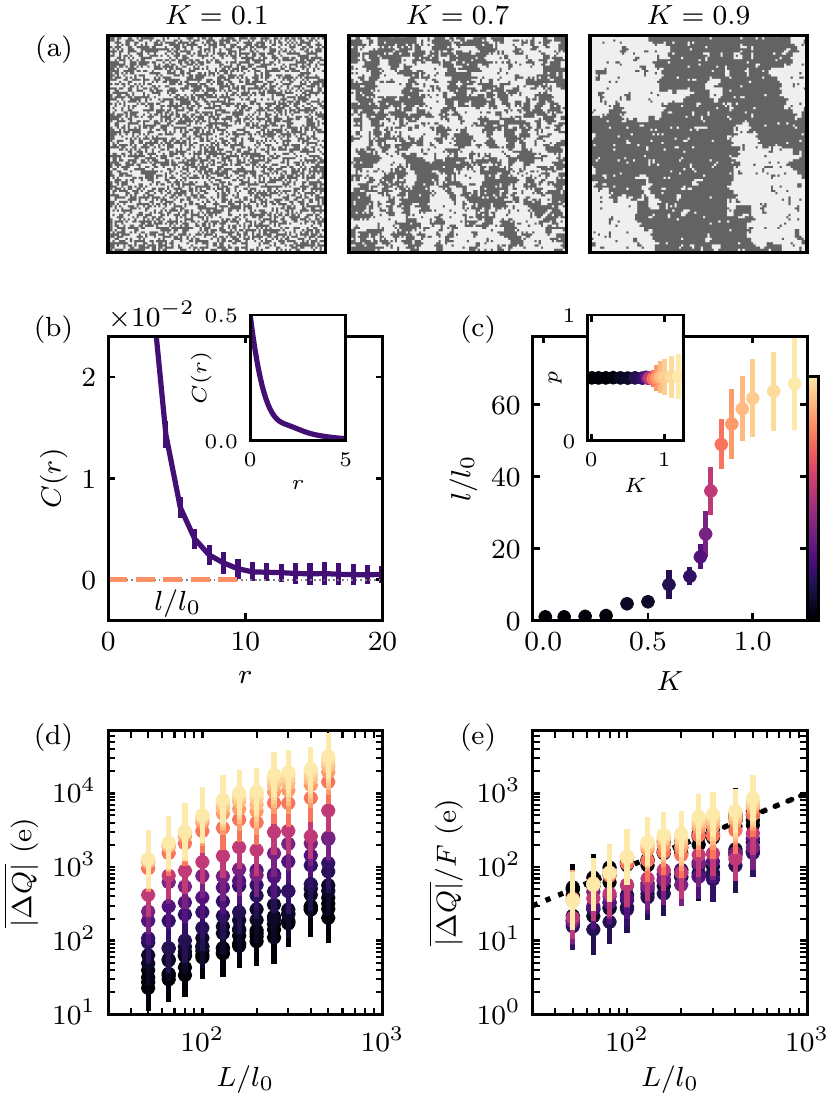} 
\caption{\label{fig4} (a) A generic nucleation process can lead to donor/acceptor regions with a characteristic size $l\!>\!l_0$.
(b) We measure $l/l_0$ using the radial correlation function $C(r)$ (Eq.~\ref{eq:correl}), averaged over several surfaces (error bars represent the standard deviation). The inset shows the whole vertical range.
(c) We vary $l/l_0$ through parameter $K$ in Eq.~\ref{eq:proba}. The error is calculated from the error on $C(r)$. We keep the same color scale for $l/l_0$ in all subfigures. The inset shows that $p$ remains constant.
(d) As before, the charge transferred after a contact is amplified by the introduction of $l$. The error bars indicate the standard deviation.
(e) We verify that Eq.~\ref{eq:mosaicScale} is still valid. The dotted line is the identity line.  Deviations at intermediate $K$ are due to the presence of a spectrum of feature sizes (see Supplemental Material \cite{SupplMat}).}
\end{figure}

Starting with an initial arrangement, a surface evolves until it reaches a dynamic equilibrium set by the parameters in Eq.~\ref{eq:proba} (for a movie, see the Supplemental Material~\cite{SupplMat}).  Three examples for different $K$ are shown in Fig.~\ref{fig4}a .  To characterize these surfaces in the context of our analysis, we measure $p$ and $l/l_0$.  Determining $p$ is trivial. To get $l/l_0$ we start by calculating the correlation function,
\begin{equation}
    C(r) = \langle s(R) s(R+r) \rangle - \langle s(R) \rangle\langle s(R+r) \rangle,
    \label{eq:correl}
\end{equation}
where $s$ is the site identity at a position $R$, $r$ is the distance from the point $R$, and averages denoted by $\langle$ $\rangle$ are over all sites separated by $r$. Figure~\ref{fig4}b shows an example of $C(r)$ for a particular surface. We use the first zero crossing (within the standard deviation of $C(r)$) to define the correlation length, $l$, as it corresponds to the typical distance before having an equal probability of switching from donor to acceptor (or \textit{vice versa}).  In the Supplemental Material, we show that this identically recovers the input correlation lengths in of our other set of simulations \cite{SupplMat}.  In Figure~\ref{fig4}c, we show how sweeping through the parameter $K$ ($P_0$ fixed) allows us to explore nearly two orders of magnitude in $l/l_0$ (with $p\!\approx\!0.5$).

After they have reached a dynamic equilibrium, we freeze these surfaces and then use them as input for contact experiments, with the same transfer rules as before.  We generate 20 surfaces for various combinations of correlation length $l/l_0$ and system size $L/l_0$, and calculate $\overline{|\Delta Q|}$ for every permutation (Fig.~\ref{fig4}d). The effect of spatial correlations in this physically-derived system is the same as with the synthetic surfaces, \textit{i.e.}~the magnitude of transfer increases with the correlation length. Figure~\ref{fig4}e presents the results rescaled using Eq.~\ref{eq:mosaicScale}, which largely collapses them onto a single line with the predicted unity slope.  In the Supplemental Material \cite{SupplMat}, we show that the deviation for intermediate $K$ is due a broader `spectrum' of length scales for $l$ than in our synthetic simulations. 

Based on widely held assumptions and consideration of existing experiments, we have developed a quantitative, scale-spanning model for same-material tribocharging.  We have shown that the intermediate scale corresponding to donor/acceptor spatial correlations, $l$, plays a crucial role, amplifying the amount of charge transferred in a single contact, and suppressing otherwise overwhelming fluctuations in sequential contacts.  We have furthermore introduced a candidate mechanism for how the intermediate length scale emerges, which is based on the energetics of donor/acceptor interactions during the formation process.  Our results allow us to quantitatively resolve the inconsistency encountered by Apodaca~\cite{Apodaca:2009dr}: using our analysis with their data and Baytekin's correlation length~\cite{Baytekin:2011bx}, we infer an elementary site size $l_0 \!\approx\! 4$~\r{A}---precisely on the atomic scale~\cite{SupplMat}.  Furthermore, our work has implications regarding the carrier and mechanism.  By elevating the Apodaca framework from a `toy model' to a `quantitative model,' we propose that confident extractions of $\alpha$ from experimental data can be made, giving information directly related to underlying atomic-scale mechanism~\cite{Kok:2009gn}.  Although our model is restricted to surfaces where the macroscopic contact area is known, these considerations become even more important in systems where roughness or stiffness play a role.

This project has received funding from the European Union’s Horizon 2020 research and innovation programme under the Marie Sk\l{}odowska-Curie Grant Agreement No. 754411.


\begin{thebibliography}{30}

\bibitem{Lacks:2019dx}  D.J.~Lacks and T.~Shinbrot, Nat.~Rev.~Chem.~\textbf{3}, 465-476 (2019).

\bibitem{Steinpilz:2020iy} T.~Steinpilz, K.~Joeris, F.~Jungmann, D.~Wolf, L.~Brendel, J.~Teiser, T.~Shinbrot and G.~Wurm, Nat.~Phys.~\textbf{16}, 225-229 (2020).

\bibitem{Desch:2000} S.J.~Desch and J.N.~Cuzzi, Icarus~\textbf{143}, 87-105 (2000).

\bibitem{Wurm:2019}  G.~Wurm, L.~Schmidt, T.~Steinpilz, L.~Boden and J.~Teiser, Icarus~\textbf{331}, 103-109 (2019).

\bibitem{Schrader:2018eh} D.L.~Schrader, K.~Nagashima, S.R.~Waitukaitis, J.~Davidson, T.J.~McCoy, H.C.~Connolly Jr., and D.S.~Lauretta, Geochim.~Cosmochim.~Ac.~\textbf{223}, 405-421 (2018).

\bibitem{Berdeklis:2001ta}  P.~Berdeklis and R.~List, J.~Am.~Met.~So.~\textbf{58}, 2751-2770 (2001).

\bibitem{Baytekin:2013dq} H.T.~Baytekin, B.~Baytekin, T.M.~Hermans, B.~Kowalczyk and B.A.~Grzybowski, Science~\textbf{341}, 1368-1371 (2013).

\bibitem{Abbasi:2007cv} T.~Abbasi and S.A.~Abbasi, J.~Hazard.~Mat.~\textbf{140} 7-44 (2007).

\bibitem{Kanik:2014bn} M.~Kanik, O.~Aktas, H.S.~Sen, E.~Durgun and M.~Bayindir, ACS~Nano~\textbf{8}, 9311-9323 (2014).

\bibitem{Musa:2018uy}  U.G.~Musa, S.D.~Cezan, B.~Baytekin and H.T.~Baytekin, Sci.~Rep.~\textbf{8}, 2472 (2018).

\bibitem{Wang:2013bq} Z.L.~Wang, ACS~Nano~\textbf{7} 9533-9557 (2013).

\bibitem{Hu:2012jf}  W.~Hu, L.~Xie and X.~Zheng, Appl.~Phys.~Lett.~\textbf{101}, 114107 (2012).

\bibitem{Waitukaitis:2013fwa}  S.R.~Waitukaitis and H.M.~Jaeger, Rev.~Sci.~Instrum.~\textbf{84}, 025104 (2013).

\bibitem{Waitukaitis:2014cg}  S.R.~Waitukaitis, V.~Lee, J.M.~Pierson, S.L.~Forman and H.M.~Jaeger, Phys.~Rev.~Lett.~\textbf{112}, 218001 (2014).

\bibitem{Collins:2018} A.~L.~Collins, C.~G.~Camara, E.~V.~Van~Cleve and S.~J.~Putterman, Rev.~Sci.~Instrum.~\textbf{89}, 013901 (2018).

\bibitem{Lee:2018gg} V.~Lee, N.M.~James, S.R.~Waitukaitis and H.M.~Jaeger, Phys.~Rev.~Mat.~\textbf{2}, 035602 (2018).

\bibitem{Harris:2019} I.A.~Harris, M.X.~Lim and H.M.~Jaeger, Phys.~Rev.~Mat.~\textbf{3}, 085603 (2019).

\bibitem{Lacks:2011hm}  D.J.~Lacks and M.R.~Sankaran, J.~Phys.~D: Appl.~Phys.~\textbf{44}, 453001 (2011).





\bibitem{Lowell:2000jk}  J.~Lowell and W.S.~Truscott, J.~Phys.~D: Appl.~Phys.~\textbf{19}, 1273-1280 (1986).

\bibitem{Forward:2009in}  K.M.~Forward, D.J.~Lacks and R.M.~Sankaran, Geophys.~Res.~Lett.~\textbf{36}, L13201 (2009).


\bibitem{Lowell:2000bl}  J.~Lowell and W.~S.~Truscott, J.~Phys.~D: Appl.~Phys.~\textbf{19}, 1281 (1986).

\bibitem{Lacks:2007ge} D.~J.~Lacks, N.~Duff and S.~K.~Kumar, Phys.~Rev.~Lett.~\textbf{100}, 188305 (2008).

\bibitem{Lacks:2008hz}  D.~J.~Lacks and A.~Levandovsky, J.~Electrostat.~\textbf{65}, 107-112 (2007).

\bibitem{Duff:2008} N.~Duff and D.J.~Lacks, J.~Electrostat.~\textbf{66}, 51-57 (2008).

\bibitem{Lacks:2016} D.J.~Lacks and R.M.~Sankaran, Particul.~Sci.~Technol.~\textbf{34}, 55-62 (2016).

\bibitem{Shinbrot:2017} T.~Shinbrot, M.~Rutala and H.~Herrmann, Phys.~Rev.~E~\textbf{96}, 032912 (2017).

\bibitem{Siu:2014} T.~Siu, J.~Cotton, G.~Mattson and T.~Shinbrot, Phys.~Rev.~E~\textbf{89}, 052208 (2014).

\bibitem{Kolehmainen:2018} J.~Kolehmainen, A.~Ozel, Y.~Gu, T.~Shinbrot and S.~Sundaresan, Phys.~Rev.~Lett.~\textbf{121}, 124503 (2018).

\bibitem{Yoshimatsu:2016} R.~Yoshimatsu, N.A.M.~Ara\'ujo, G.~Wurm, H.J.~Herrmann and T~Shinbrot, Sci.~Rep.~\textbf{7}, 39996 (2016).

\bibitem{Yoshimatsu:2016b} R.~Yoshimatsu, N.A.M.~Ara\'ujo, T.~Shinbrot and H.J.~Herrmann, Soft~Mat.~\textbf{12}, 6261–6267 (2016).

\bibitem{Sow:2012} M.~Sow, D.J.~Lacks and M.R.~Sankaran, J.~Appl.~Phys.~\textbf{112}, 084909 (2012). 

\bibitem{Sow:2012b} M.~Sow, R.~Widenor, A.~Kumar, S.W.~Lee, D.J.~Lacks and R.M.~Sankaran, Angew.~Chem.~Int.~Ed.~\textbf{51}, 2695 (2012).

\bibitem{Sow:2013} M.~Sow, D.J.~Lacks and R.M.~Sankaran, J.~Electrostat.~\textbf{71}, 396–399 (2013).

\bibitem{Apodaca:2009dr} M.M.~Apodaca, P.J.~Wesson, K.J.M.~Bishop, M.A.~Ratner and B.A.~Grzybowski, Angew.~Chem.~Int.~Ed.~\textbf{49}, 946-949 (2009).

\bibitem{Baytekin:2011bx} H.~T.~Baytekin, A.~Z.~Patashinski, M.~Branicki, B.~Baytekin, S.~Soh and B.~A.~Grzybowski, Science~\textbf{333}, 308-312 (2011).

\bibitem[{Sup()}]{SupplMat}
\bibinfo{title}{{See Supplemental Material at [URL will be inserted by publisher] for additional details.}}

\bibitem{Haberle:2019}  J.~Haeberle, A.~Schella, M.~Sperl, M.~Schröter and P.~Born, Soft Matter \textbf{14}, 4987-4995 (2019).

\bibitem{Wang:2017} A.~E.~Wang, P.~S.~Gil, M.~Holonga, Z.~Yavuz, H.~T.~Baytekin, R.~M.~Sankaran and D.~J~.Lacks, Phys.~Rev.~Mater.  \textbf{1}, 035605 (2017).

\bibitem{Xie:2016gr}  L.~Xie, N.~Bao, Y.~Jiang and J.~Zhou, AIP Adv.~\textbf{6}, 035117 (2016).

\bibitem{Yu:2017} H.~Yu, L.~Mu and L.~Xie, J.~Electrostat.~\textbf{90}, 113-122 (2017).

\bibitem{Brune:1998} H.~Brune, Surf.~Sci.~Rep. \textbf{31}, 121-229 (1998).

\bibitem{Varanasi:2009} K.~K.~Varanasi, M.~Hsu, N.~Bhate, W.~Yang and T.~Deng, Appl.~Phys.~Lett. \textbf{95}, 094101 (2009).

\bibitem{Lu:2018} B.~L\"u, G.~A.~Almyras, V.~Gervilla, J.~E.~Greene and K.~Sarakinos, Phys.~Rev.~Mater. \textbf{2}, 063401 (2018).



\bibitem{Kok:2009gn} J.F.~Kok and D.J.~Lacks, Phys.~Rev.~E \textbf{79}, 051304 (2009).



\end{thebibliography}
\end{document}


\preprint{APS/SI-correlation-0.1}

\title{A quantitatively consistent, scale-spanning model for same-material tribocharging:\\ supplemental information}

\author{Galien Grosjean}%
 \email{galienmariep.grosjean@ist.ac.at}
\author{Sebastian Wald}
\author{Juan Carlos Sobarzo}
\author{Scott Waitukaitis}
\affiliation{%
 IST Austria\\
 Lab Building West\\
 Am Campus 1\\
 3400 Klosterneuburg AT
}%

\date{\today}

\maketitle

\section{Synthetic surfaces}
\label{SI:surface}

As discussed in the main text, our numerical simulations used in Figs.~1-3 of the main paper involve making pairs of `synthetic' surfaces, \textit{i.e.}~matrices, whose elements are assigned as either donor (1) or acceptor (0).  We call these `synthetic' as the process by which we create them is not physically derived, but rather comes from a numerical algorithm that allows us to efficiently produce random surfaces with a desired donor probability $p$ and legnthscale $l/l_0$.  Later in this supplement, we will discuss the `physical' algorithm based on nucleation, which applies to Fig.~4 of the main paper.  

\begin{figure}
    \centering
    \subfigure{\includegraphics[width=14cm]{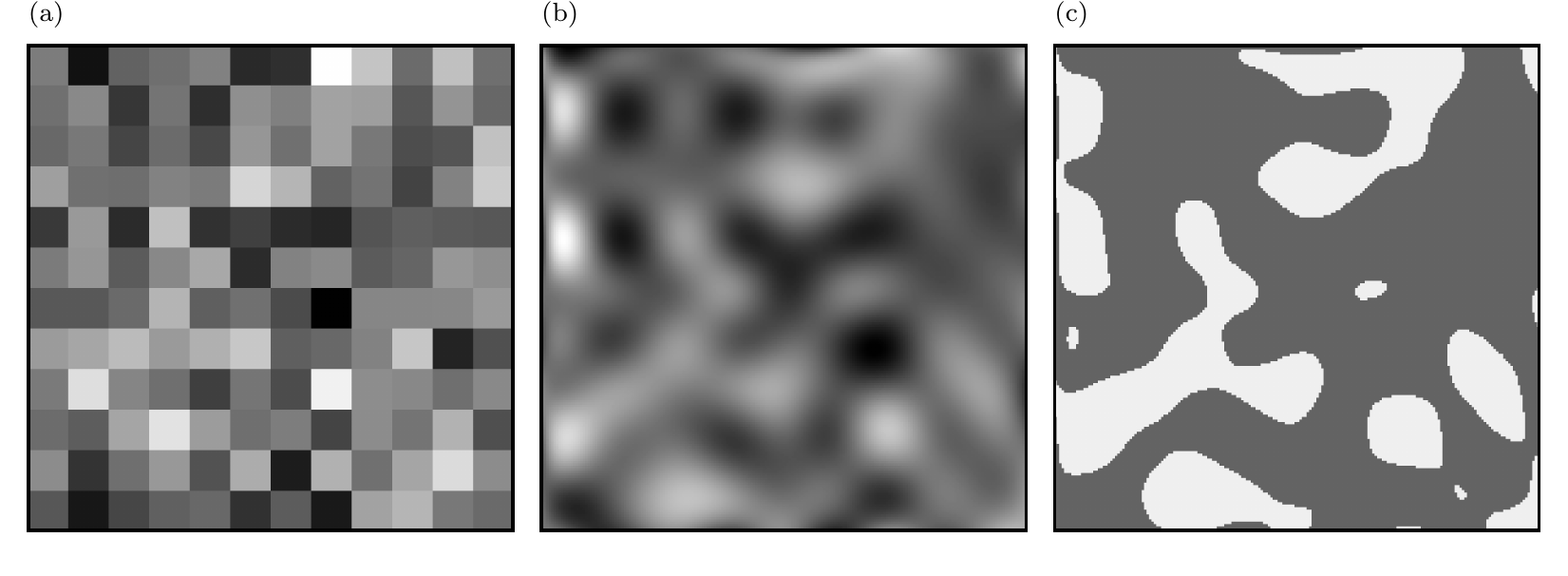}}
    \caption{Successive steps in creating a surface of donor and acceptor sites. (a) A coarse square matrix of size $\sim L/l \times L/l$ is filled with normally distributed random values. (b) This is then used to populate every $l/l_0$\ts{th} point of the main $L/l_0 \times L/l_0$ matrix; the elements in between (at the scale $l_0$) are filled by interpolation. (c) Finally, this smooth field is binarized such that the probability that an element has a value above a calculated threshold is on average equal to the desired donor probability, $p$. Donors are shown here in dark gray, and acceptors in light gray. The parameters are $L/l_0=200$, $l/l_0=20$ and $p=0.25$. 
    }
    \label{SI:fig:matrixConstruction}
    \centering
    \includegraphics[]{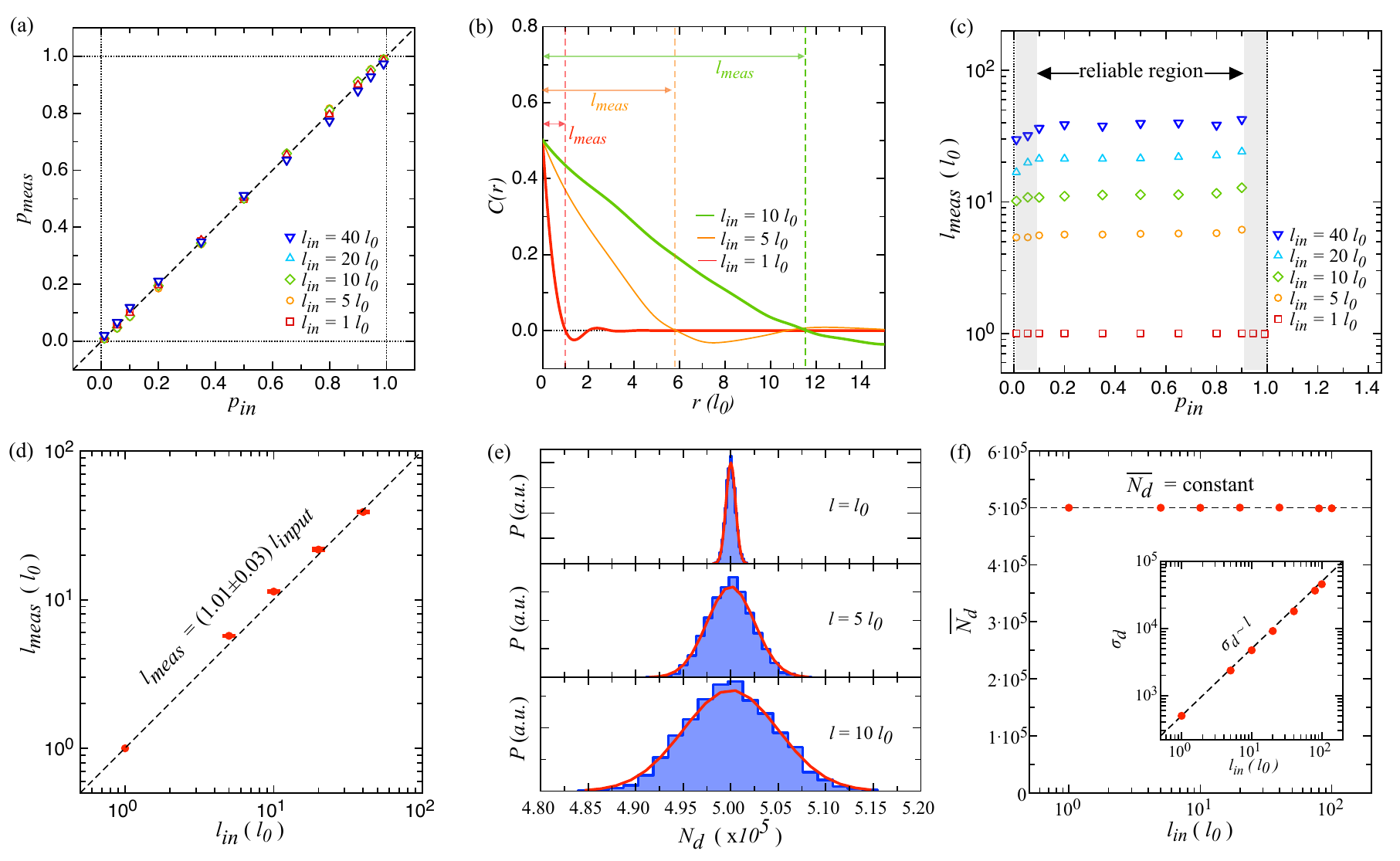}
    \caption{Tests for synthetic surfaces.  (a) Measured donor probabilities, $p_{meas}$, \textit{vs.}~input values, $p_{in}$ for a range of input length scales.  Each point represents the average value, $p_{meas}=\overline{N_d}/N$, for 30 generated surfaces, each with $L/l_0=300$ ($N=300\times300$).  (b)  Average measured correlation function, $C(r)$, for surfaces with $l_{in}=1 l_0$ (red), $5l_0$ (orange), and $10l_0$ (green).  From each of these curves, we extract $l_{meas}$ as the position $r$ of the first zero crossing in $C(r)$ (vertical dashed lines).  (c) Measured length scale, $l_{meas}$, \textit{vs.}~input probability for the same surfaces as in (a).  For $0.1<p_{in}<0.9$, $l_{meas}$ is stable and approximately equal to $l_{in}$.  (d) Average values of $l_{meas}$ in the reliable region from (c) \textit{vs.}~$l_{in}$, which reveals they are approximately equal.  (e) Distributions in the number of generated donors, $N_d$, for 5000 instances of a $1000\times1000$ element surface with $p=0.5$ and $l/l_0=1$, 5, and 10.  The red curves are Gaussian fits.  (e)  While the mean value $\overline{N_d}$ (and hence probability $p$) is constant as a function of $l$, the distribution widths $\sigma_d$ are proportional to $l$ (inset).  Given $p_{meas}=p_{in}$ and $l_{meas}=l_{in}$, we simply refer to them as $p$ and $l$, respectively.
    }
    \label{SI:fig:PLsweeps}
\end{figure}

The input parameters to construct the synthetic surfaces are the numbers of rows and columns, $L/l_0$, the typical size of the donor and acceptor patches, $l/l_0$, and the donor probability $p$. The size $l_0$ corresponds to a single matrix element in our simulations.  To fill our matrices, we first create a coarse matrix of size $L/l \times L/l$ and fill it with random values drawn from a normal distribution centered at zero and with standard deviation one (Suppl.~Fig.~\ref{SI:fig:matrixConstruction}a; if $L/l$ is not an integer we round up).  We use these values to fill every ($l/l_0$)\ts{th} element of the final matrix, \textit{e.g.}~if $l/l_0=10$ then we fill points $[i,j]=[0,0], [10,0], [20,0],\hdots, [10,0], [10,10],\hdots$ \textit{etc.}  We assign values to the matrix elements between by performing a 5\ts{th} order 2-dimensional interpolation (Suppl.~Fig.~\ref{SI:fig:matrixConstruction}b).  Next, we threshold this smoothly varying field to make a binary field. The value of the threshold, $T$, depends on the probability $p$, and is determined by the equation
\begin{equation}
    p = \frac{1}{\sqrt{2\pi}}\int_{-\infty}^T \exp\left(\frac{-x^2}{2}\right)dx.
\end{equation}
After solving this equation numerically, we assign all matrix elements with values less than $T$ to be 1 (donor), and all with values higher to be 0 (acceptor).  This results in a final binary matrix, as shown in Suppl.~Fig.~\ref{SI:fig:matrixConstruction}c or in Fig.~1 of the main paper.  Qualitatively, these surfaces have heterogeneous features reminiscent of those seen by KPFM in Ref.~\cite{SI:Baytekin:2011bx}.

We perform several tests to confirm that our procedure creates surfaces with the characteristics we desire.  First, in Suppl.~Fig.~\ref{SI:fig:PLsweeps}a, we show that the measured probability for being a donor (\textit{i.e.}~$p_{meas} = \overline{N_d}/N$ for several hundred generated matrices) is approximately equal to the input probability, $p_{in}$.  We therefore can simply write $p_{meas}=p_{in}\equiv p$.  To ensure the length scale of the generated donor/acceptor features matches our input, $l_{in}$, we measure the radial correlation function, $C(r)$, exactly as prescribed in the main text in the context of our nucleation simulations, \textit{i.e.}, 
\begin{equation}
     C(r) = \langle s(R) s(R+r) \rangle - \langle s(R) \rangle\langle s(R+r) \rangle,
\end{equation}
where $s$ is the site identity (0 for acceptors and 1 for donors), $R$ is the position of an element, $r$ is the distance away from that element, and the averages $\langle$ $\rangle$ are done over all positions separated by a distance $r$.  Suppl.~Fig.~\ref{SI:fig:PLsweeps}b shows $C(r)$ for a few values of $l_{in}$ where, as mentioned in the main paper, we take the zero crossing to define $l_{meas}$.  In Suppl.~Fig.~\ref{SI:fig:PLsweeps}c we plot $l_{meas}$ as a function of $p_{in}$.  This plot reveals that the measured value, $l_{meas}$, is stable for intermediate values of $p_{in}$, but exhibits deviations for  $p_{in}>0.9$ and $p_{in}<0.1$; we therefore restrict our work to this range.  In Suppl.~Fig.~\ref{SI:fig:PLsweeps}d, we plot the averages over the stable region to show that $l_{meas}=(1.01 \pm 0.03) l_{in}$.  We therefore can simply write $l_{meas}=l_{in}\equiv l$.  This validates that our `synthetic' protocol allows us to create random surfaces with features of a desired size.  

Finally, we make a brief aside to look at the behavior of the number of donors, $N_d$, as a function of the length scale $l/l_0$.  Supplementary Fig.~\ref{SI:fig:PLsweeps}e shows distributions in $N_d$ for different $l/l_0$.  As can be seen, the mean value of $N_d$ (or equivalently the probability for a site to be a donor, $p$) does not change with $l$ (Suppl.~Fig.~\ref{SI:fig:PLsweeps}f).  However, the widths of these distributions \textit{do} change with $l$ (inset of Suppl.~Fig.~\ref{SI:fig:PLsweeps}f).  As mentioned in the main paper, such broadening turns out to be the critical ingredient in how the introduction of the intermediate length scale $l$ leads to charge-transfer enhancement.  

\section{Theory}
\label{SI:Theory}

\subsection{First contact when $l=l_0$}

We begin by determining the distribution in the total charge transfer we expect in a first contact \textit{solely} from the left surface to the right surface, $P(Q_\shortrightarrow)$.  Considering a given site pair $[i,j]$, there are two relevant and mutually exclusive outcomes:  (1) left-to-right transfer does occur, which has probability $p(1-p)\alpha$, and (2) left-to-right doesn't occur, which has probability $1-p(1-p)\alpha$ (see Suppl.~Table~\ref{SI:indepProb}).  The probability for transfer at any one site pair is independent from all others.  Thus, for large $L/l_0$, the probability distribution for total left-to-right transfer is Gaussian with mean $e(L/l_0)^2p(1-p)\alpha$ and width $eL/l_0\sqrt{p(1-p)\alpha(1-p(1-p)\alpha)}$, \textit{i.e.},
\begin{equation}
    P(Q_\shortrightarrow) = \frac{1}{eL/l_0\sqrt{2\pi p(1-p)\alpha(1-p(1-p)\alpha)}}\exp \left(\frac{(Q_\shortrightarrow-eL^2/l_0^2p(1-p)\alpha)^2}{2e^2L^2/l_0^2p(1-p)\alpha(1-p(1-p)\alpha)}\right),
\end{equation}
If we had instead begun by considering right-to-left transfer, we would have found similar probabilities (second part of Suppl.~Table~\ref{SI:indepProb}) and a corresponding distribution, $P(Q_\shortleftarrow)$.

Strictly speaking, the expressions for $P(Q_\shortrightarrow)$ and $P(Q_\shortleftarrow)$ are only valid when considering left-to-right and right-to-left transfer independently.  This is because, at a given site pair, left-to-right and right-to-left transfer cannot occur simultaneously.  The exact way to handle this is to consider the situation where three mutually exclusive outcomes are possible:  (1) right-to-left transfer occurs, (2) left-to-right transfer occurs, or (3) neither right-to-left nor left-to-right transfer occurs.  These outcomes and their corresponding probabilities are listed in Suppl.~Table~\ref{SI:coupledProb}.  Rather than solving this three-outcome problem exactly, we make progress by assuming that left-to-right transfer and right-to-left transfer are independent.  This leads to four outcomes, with probabilities as in the column `relaxed prob.' in Suppl.~Table~\ref{SI:coupledProb}.  In this case, the probabilities are products of those in Suppl.~Table~\ref{SI:indepProb}, and all involve terms on the order of $1$, $p(1-p)\alpha$, or $(p(1-p)\alpha)^2$.  At its maximal value, which occurs when $p=0.5$ and $\alpha=1$, the squared term is 1/4\ts{th} as large as the first order term.  However, the experiments by Apodaca \cite{SI:Apodaca:2009dr} suggest $\alpha$ is much smaller than one (more on the order of 0.1), thus realistically we expect the second order term to be $\lesssim1/40$\ts{th} as large as the first order term.  This means that we can ignore the squared terms, which renders the relaxed case approximately equal to the strict case (column `approx.~relaxed prob.' in Suppl.~Table~\ref{SI:coupledProb}).

\begin{table}[t]
\begin{center}
\begin{tabular}{ll}
\begin{tabular}{ |c|c|} \hline
\multicolumn{2}{|c|}{\textbf{Left-to-right}} \\ \hline
\textbf{outcome} & \textbf{probability}  \\ \hline
transfer $\rightarrow$ & $p(1-p)\alpha$ \\ \hline
no transfer $\rightarrow$ & $1-p(1-p)\alpha$ \\
\hline
\end{tabular}
\hspace*{3em}
\begin{tabular}{|c|c|} \hline
\multicolumn{2}{|c|}{\textbf{Right-to-left}} \\ \hline
\textbf{outcome} & \textbf{probability}  \\ \hline
transfer $\leftarrow$ & $p(1-p)\alpha$ \\ \hline
no transfer $\leftarrow$ & $1-p(1-p)\alpha$ \\
\hline
\end{tabular}
\end{tabular}
\caption{Outcomes and probabilities for left-to-right and right-to-left transfer if each case is considered separately.  Strictly speaking, the tables can only be considered `one at a time' as the left-to-right and right-to-left outcomes at a given site-pair cannot occur simultaneously.\\~\\}
\label{SI:indepProb}
\begin{tabular}{|c|c|c|c|}  \hline
\textbf{outcome} &  \textbf{strict prob.} & \textbf{relaxed prob.} & \textbf{approx.~relaxed prob.}  \\ \hline
transfer $\rightarrow$ & $p(1-p)\alpha$ & $p(1-p)\alpha(1-p(1-p)\alpha)$ & $\sim p(1-p)\alpha$\\ \hline
transfer $\leftarrow$ &  $p(1-p)\alpha$ & $p(1-p)\alpha(1-p(1-p)\alpha)$ & $\sim p(1-p)\alpha$ \\ \hline
transfer $\rightleftarrows$ & 0 &  $(p(1-p)\alpha)^2$ & $\sim0$ \\ \hline 
no transfer $\rightleftarrows$ & $1-2p(1-p)\alpha$ & $(1-p(1-p)\alpha)^2$ &  $\sim1-2p(1-p)\alpha$   \\ \hline
\end{tabular}
\caption{Coupled outcomes and probabilities for combined left-to-right and right-to-left transfer.  The `strict prob.'~column gives the exact probabilities; with three outcomes, it requires multinomial analysis.  The `relaxed prob.'~column gives the coupled probabilities that arise if right-to-left transfer and left-to-right transfer (as in Suppl.~Table~\ref{SI:indepProb}) are assumed independent.  Ignoring terms of the order $(p(1-p)\alpha)^2$ leads to the `approx.~relaxed prob.'~column, which is nearly equal to the strict case.}
\label{SI:coupledProb}
\end{center}
\end{table}

\begin{figure}
    \centering
    \includegraphics[]{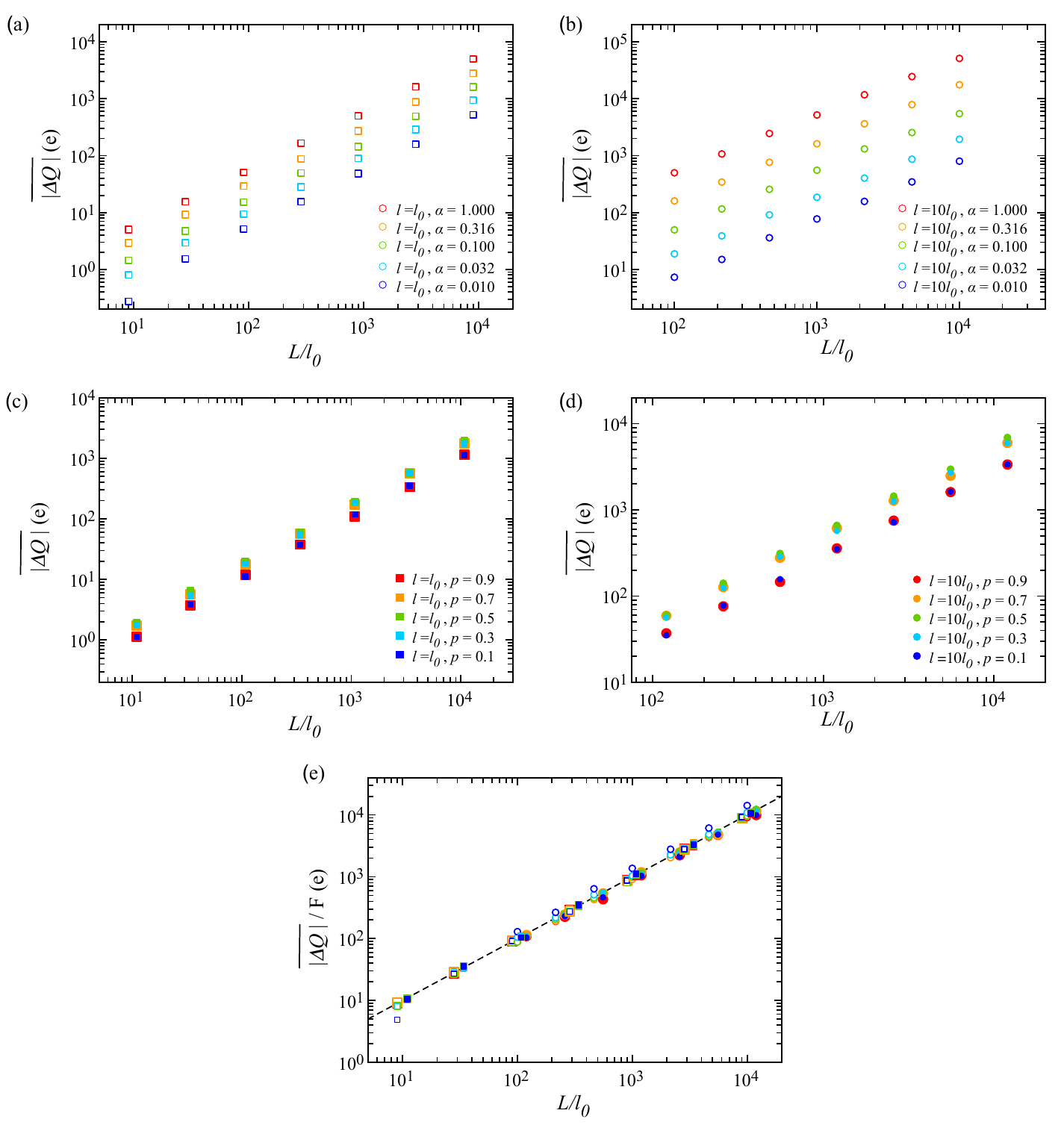}
    \caption{Variation in charge transfer with $p$ and $\alpha$.  (a) Mean absolute value in charge transferred, $\overline{|\Delta Q|}$, \textit{vs.}~$L/l_0$ for several different values of $\alpha$ and with $l=l_0$ (symbols indicated in plot legend).  (b) $\overline{|\Delta Q|}$ \textit{vs.}~$L/l_0$ for several different values of $\alpha$ and with $l=10l_0$.  (c) $\overline{|\Delta Q|}$ \textit{vs.}~$L/l_0$ for several different values of $p$ and with $l=l_0$.  (d) $\overline{|\Delta Q|}$ \textit{vs.}~$L/l_0$ for several different values of $p$ and with $l=10 l_0$.  (e) Collapsed master curve from all data in plots (a)-(d).  The rescaling factor is $F=\sqrt{4p(1-p)\alpha/\pi}$ when $l=l_0$, and $F=\alpha l/l_0 \sqrt{4p(1-p)/\pi}$ when $l=10 l_0$.  In panels (a) and (b), $p=0.5$; in panels (c) and (d), $\alpha=0.1$.  The number of independent surface pairs generated to calculate the mean for each point is 500.
    }
    \label{SI:fig:PAsweeps}
\end{figure}

The preceding paragraph reveals that, to a good approximation, we can treat the distributions $P(Q_\shortrightarrow)$ and $P(Q_\shortleftarrow)$ as independent. Since they are both Gaussian distributed, so too is $P(\Delta Q)$, but with mean $\overline{\Delta Q}=\overline{Q_\shortrightarrow}-\overline{Q_\shortleftarrow}=0$ and width $\sigma=\sqrt{\sigma_\shortrightarrow^2 + \sigma_\shortleftarrow^2}$.  If we again neglect terms on the order of $(p(1-p)\alpha)^2$, this leads to the $l=l_0$ probability distribution,
\begin{equation}
    P(\Delta Q) = \frac{1}{eL/l_0\sqrt{4\pi p (1-p) \alpha}}\exp \left(-\frac{\Delta Q^2}{4e^2L^2/l_0^2 p(1-p)\alpha}\right).
\label{SI:eq:smallestScale}
\end{equation}
As the mean value of $\Delta Q$ is zero, $\overline{|\Delta Q|}$ is necessarily proportional to the distribution width.  The exact expression is
\begin{equation}
    \overline{|\Delta Q|} = \sqrt{\frac{2}{\pi}}\sigma = \sqrt{\frac{2}{\pi}}\frac{eL}{l_0}\sqrt{2p(1-p)\alpha}.
    \label{SI:eq:l_eq_l0}
\end{equation}
This result therefore reproduces the $\overline{|\Delta Q|}\propto L\propto\sqrt{A}$ scaling found previously \cite{SI:Apodaca:2009dr}, except for the square root dependency in $\alpha$.  In Suppl.~Fig.~\ref{SI:fig:PAsweeps}a and~\ref{SI:fig:PAsweeps}c, we run multiple simulations to plot $\overline{|\Delta Q|}$ \textit{vs.}~$L/l_0$ for several values of $\alpha$ and $p$. In Suppl.~Fig.~\ref{SI:fig:PAsweeps}e, we show that rescaling by the appropriate factor, $F=\sqrt{4p(1-p)\alpha/\pi}$, collapses all of our simulated numerical data for $l=l_0$ onto a single master curve. 

\subsection{First contact when $l>>l_0$}

In the analysis for $l=l_0$, we started by considering the total left-to-right charge transfer, $Q_\shortrightarrow$.  A fundamental assumption was that the probability for transfer at any site pair was independent from all others.  When patches are present, we can no longer make this assumption.  This is because if a given site is a donor (acceptor), it is now more likely that its neighbor is also a donor (acceptor).  We account for this by imagining a rescaled matrix consisting of larger patches, each patch containing $n=(l/l_0)^2$ elementary sites. There is therefore a total of $N'=N/n$ such patches. We randomly assign the patches of this rescaled matrix as donors/acceptors (still with probabilities $p$ and $1-p$) to properly account for correlations.  If a patch is assigned as a donor (acceptor), then all $n$ elementary sites within it are donors (acceptors). Critically, we will assume that the transfer of charge still occurs independently for each site with probability $\alpha$.

We now consider how many donor patches on the left surface face acceptors the right, $N'_\shortrightarrow$ (holding off on the effect of the transfer probability, $\alpha$, until later).  The probability that a donor patch faces an acceptor is given by $p(1-p)$, thus we can expect $P(N'_\shortrightarrow)$ to be Gaussian with mean $L^2/l^2p(1-p)$ and width $L/l\sqrt{p(1-p)(1-p(1-p))}$ (and similarly for $P(N'_\shortleftarrow)$).  Next, we repeat the strategy from the $l=l_0$ case and treat $N'_\shortrightarrow$ and $N'_\shortleftarrow$ as independent. The probability distribution for $N'_\shortrightarrow - N'_\shortleftarrow$ is Gaussian with zero mean and width $L/l_0\sqrt{2p(1-p)(1-p(1-p))}\approx L/l_0\sqrt{2p(1-p)}$.  For every donor patch that faces an acceptor patch, $n=(l/l_0)^2$ elementary sites interact.  If $l>>l_0$, then the mean number of transfers between patches, $\alpha n$, is much larger than the fluctuations, $\sqrt{n\alpha(1-\alpha)}$.  Thus, to a good approximation, we can expect an amount of charge, $e\alpha (l/l_0)^2$, to be transferred for \textit{every} donor patch that faces an acceptor patch, which leads us to Eq.~2 of the main text,
\begin{equation}
    P(\Delta Q) = \frac{1}{\alpha e Ll/l_0^2\sqrt{4\pi p(1-p)}}\exp \left(-\frac{\Delta Q^2}{4\alpha^2e^2L^2l^2/l_0^4p(1-p)}\right).
\label{SI:eq:mosaicScale}
\end{equation}
The width of this distribution, $\sigma=\alpha e Ll/l_0^2\sqrt{2p(1-p)}$, differs from the $l=l_0$ case in two ways.  First, its $\alpha$-dependence is linear as opposed to square root in Suppl. Eq.~\ref{SI:eq:smallestScale}. Second, it grows proportionally with the length $l$.  This ultimately derives from the increased variability in the number of donors on each surface (consistent with the simulation results in Suppl.~Fig.~\ref{SI:fig:PLsweeps}d-f).  The mean absolute charge transferred is given by
\begin{equation}
    \overline{|\Delta Q|} = \sqrt{\frac{2}{\pi}}\frac{e\alpha Ll}{l_0^2}\sqrt{2p(1-p)}.
    \label{SI:eq:l_eq_l}
\end{equation}
In Suppl.~Fig.~\ref{SI:fig:PAsweeps}b and~\ref{SI:fig:PAsweeps}d, we run many simulations to plot $\overline{|\Delta Q|}$ \textit{vs.}~$L/l_0$ for $l=10 l_0$ and several values of $\alpha$ and $p$. Rescaling by the factor, $F=\alpha l/l_0 \sqrt{4p(1-p)/\pi}$, collapses all of our numerically simulated data in the $l>>l_0$ regime (Suppl.~Fig.~\ref{SI:fig:PAsweeps}e).

\subsection{Sequential contacts: average behavior}

\begin{figure}
    \centering
    \includegraphics[]{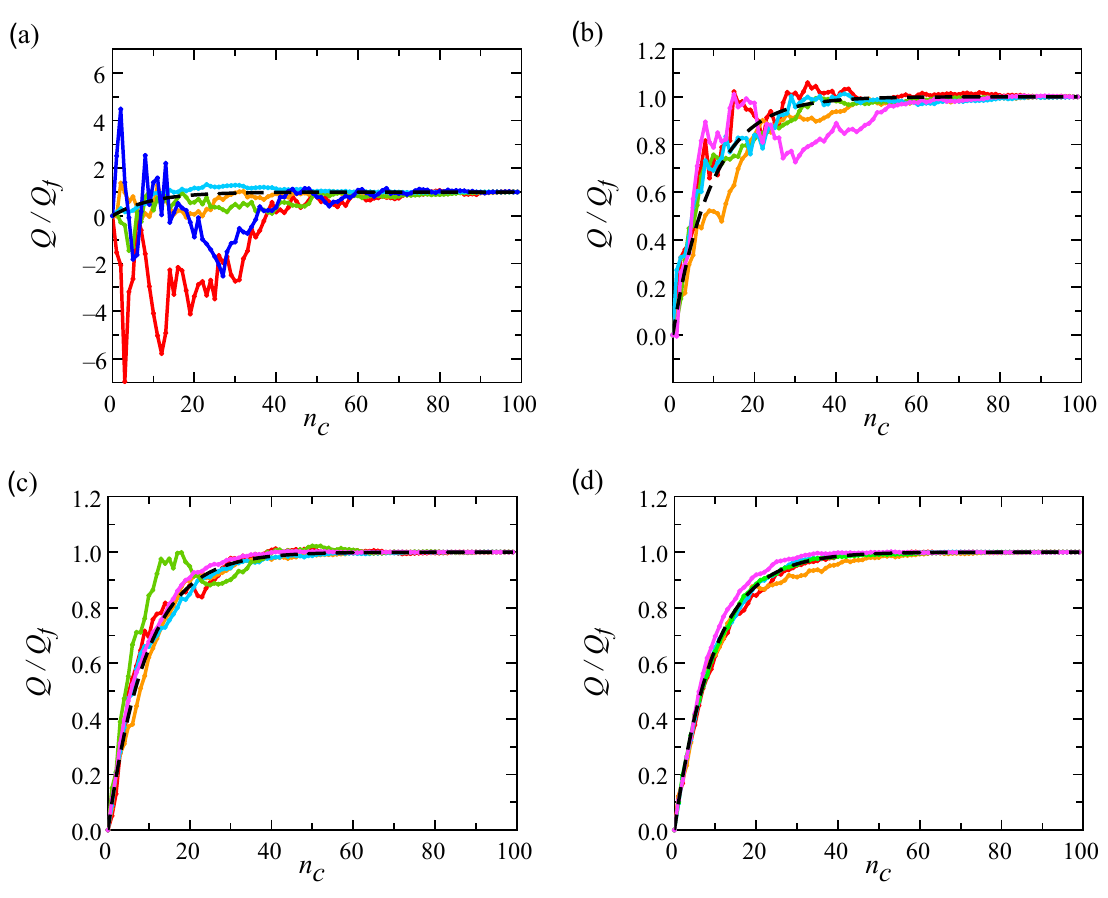}
    \caption{ Average charging behavior for sequential contacts.  (a) Charge acquired on one surface divided by its final value, $Q/Q_f$, \textit{vs.}~the number of contacts, $n_c$, for five surface pairs (different colors) with $l=l_0$, $L=1000$, $p=0.5$, and $\alpha=0.1$.  (b-d) Same as (a) but with $l=5l_0$, $l=10l_0$, and $l=20l_0$, respectively.  In all plots, the black dashed line corresponds to the prediction from Suppl.~Eq.~\ref{SI:eq:continuous3}.
    }
    \label{SI:fig:multi}
\end{figure}

Experiments show that if the same two surfaces are contacted repeatedly, charge transfer continues until one has a final total charge, $Q_f$, and the other has $-Q_f$ \cite{SI:Baytekin:2011bx}. In our numerical simulations, we see similar behavior. Suppl.~Fig.~\ref{SI:fig:multi} shows plots of $Q/Q_f$ for one surface \textit{vs.}~the number of contacts with its partner, $n_c$. As mentioned in the main text, the relative size of fluctuations on approach to the final value are larger for smaller $l$. For larger $l$, when the fluctuations are small, the curves approach a well-defined limiting curve. 

The average behavior of sequential contacts is straightforward to explain.  We again start by looking at a particular surface-pair instance, and consider the number of left-to-right and right-to-left facing donor acceptor pairs ($N_\shortrightarrow$ and $N_\shortleftarrow$, respectively).  Regardless of the size of $l$, the average amount of charge that will be transferred left-to-right in the first contact is $\alpha e(N_\shortrightarrow-N_\shortleftarrow)$.  For the subsequent contact, this reduces $N_\rightleftarrows$ by a factor $(1-\alpha)$ (on average).  Accordingly, in the second contact we expect an amount of charge transfer $\alpha e (1-\alpha)(N_\shortrightarrow-N_\shortleftarrow)$.  Continuing like this, we expect that during the $i$\ts{th} contact the amount of charge transferred is $\alpha e (1-\alpha)^{i-1} (N_\shortrightarrow-N_\shortleftarrow)$.  The total charge transferred up to contact number $n_c$ is therefore given by the sum
\begin{equation}
    Q(n_c) = \sum_{i=1}^{n_c} \alpha e (1-\alpha)^{i-1} (N_\shortrightarrow-N_\shortleftarrow).
\end{equation}
To get the limiting behavior, we evaluate the sum in the limit $n_c\rightarrow \infty$, which is given by
\begin{align}
    \lim_{n_c \to \infty} \left [ Q(n_c) \right ] & = \alpha e (N_\shortrightarrow-N_\shortleftarrow) \left[ \lim_{n_c \to \infty} \sum_{i=1}^{n_c} (1-\alpha)^{i-1} \right]  \\
    & = \alpha e (N_\shortrightarrow-N_\shortleftarrow) \left[ \frac{1}{\alpha} \right] \\
    & = e (N_\shortrightarrow-N_\shortleftarrow) \equiv Q_f.
\end{align}
Thus rescaling by $Q_f$ in any given trial, we expect that sequential contact curves should approach the geometric series
\begin{equation}
    Q(n_c)/Q_f = \sum_{i=1}^{n_c} \alpha (1-\alpha)^{i-1}.
    \label{SI:eq:multi_predict}
\end{equation}

More conveniently, when $\alpha<<1$ the charge accumulation can be approximated as a continuous expression. Assuming that the number of contacts $n_c$ is a continuous variable, an infinitesimal increase in $Q$ can be written
\begin{equation}
    \mathrm{d}Q = e \;\mathrm{d}N_{\rightleftarrows} = -\alpha N_{\rightleftarrows} e \;\mathrm{d}n_c.
    \label{SI:eq:continuous1}
\end{equation}
By integrating this equation, we find
\begin{equation}
    N_{\rightleftarrows} = N_{\rightleftarrows}^0 \exp \left(- \alpha n_c \right).
    \label{SI:eq:continuous2}
\end{equation}
The accumulated charge on the surface is given by
\begin{equation}
    Q = \int_0^Q \mathrm{d}Q = -\int_0^{n_c} \alpha e N_{\rightleftarrows}^0 \exp \left(- \alpha n_c \right) \;\mathrm{d}n_c = Q_f \left( 1 - \exp \left(-\alpha n_c \right) \right)
    \label{SI:eq:continuous3}
\end{equation}
considering that $Q_f = e N_{\rightleftarrows}^0$.
In Suppl.~Fig.~\ref{SI:fig:multi}, we plot this prediction on top of our simulated curves, which illustrates that it accurately describes the average behavior.

\subsection{Sequential contacts: fluctuations}

\begin{figure}
    \centering
    \includegraphics[]{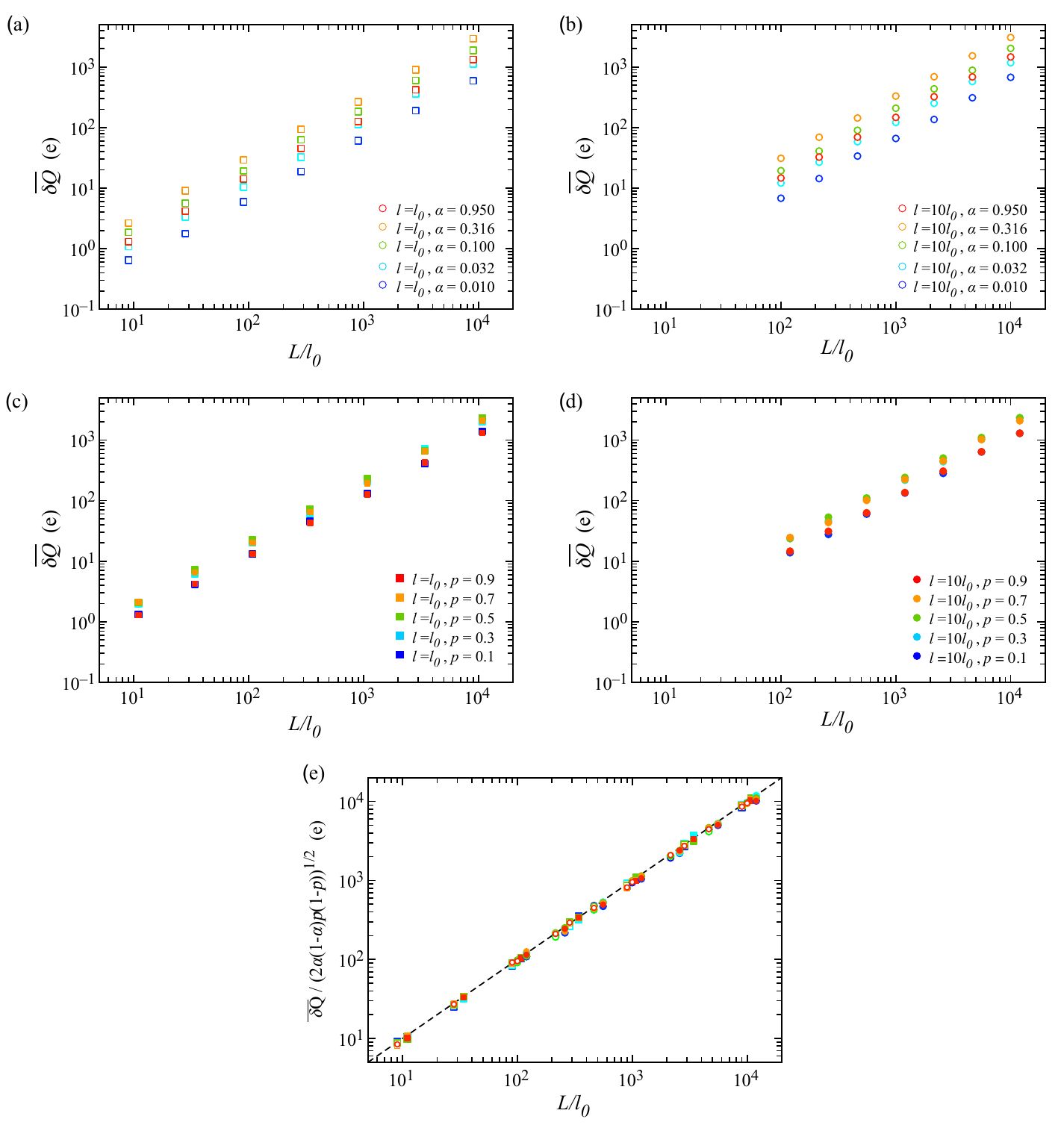}
    \caption{Multicontact fluctuations.  (a) Mean first contact fluctuations, $\overline{\delta Q}$, \textit{vs.}~$L/l_0$ for $l=l_0$ and $p=0.5$ and several different values of $\alpha$ (indicated in legend).  For each point, we calculate $\overline{\delta Q}$ as follows: (1) we create a surface pair and run 20 first contacts (resetting each time) to get $\delta Q$ for this pair, (2) we repeat for 20 surface pairs total and compute the average to get $\overline{\delta Q}$.  (b) $\overline{\delta Q}$ \textit{vs.}~$L/l_0$ for $l=10 l_0$ and $p=0.5$ and different values of $\alpha$.  (c)  $\overline{\delta Q}$ \textit{vs.}~$L/l_0$ for $l=l_0$ and $\alpha=0.1$ and different values of $p$.  (d)  $\overline{\delta Q}$ \textit{vs.}~$L/l_0$ for $l=10l_0$ and $\alpha=0.1$ and different values of $p$.  (e)  All $\overline{\delta Q}$ \textit{vs.}~$L/l_0$ rescaled by the factor $\sqrt{2\alpha(1-\alpha)p(1-p)}$, which collapses onto the predicted unity relationship. 
    }
    \label{SI:fig:FluctSweeps}
\end{figure}

Since the arrangement of facing donors/acceptors is assumed fixed, the fluctuations of charge transferred in sequential contacts must arise solely from the transfer probability, $\alpha$.  To understand how, we start by considering a specific situation with two surfaces.  We denote the number of donors on the left (right) surface that face acceptors on the right (left) as $N_{\shortrightarrow}$ ($N_{\shortleftarrow}$).  We imagine contacting the two surfaces repeatedly, each time measuring the actual number of transfers left/right as $N^*_{\rightleftarrows}$, and each time re-initializing.  The total charge transferred in one of these trials is
\begin{equation}
    \Delta Q = e(N^*_{\shortrightarrow} - N^*_{\shortleftarrow}).
\end{equation}
The mean charge transferred is $\overline{\Delta Q} = e\alpha(N_{\shortrightarrow} - N_{\shortleftarrow})$, and is in general non-zero since, in a particular instance, it is not true that $N_\shortrightarrow=N_\shortleftarrow$.  We are interested in the fluctuations about the mean, which arise due to the fluctuations ${\sigma^*_\rightleftarrows}^2=N_\rightleftarrows \alpha (1-\alpha)$ in the amounts of charge transferred right-to-left and left-to-right. The fluctuations about the mean are given by
\begin{equation}
    \delta Q = e\sqrt{ (N_\shortleftarrow+N_\shortrightarrow)\alpha(1-\alpha) }.
    \label{SI:deltaQ_1}
\end{equation}
We would now like to use this result for a particular surface-pair instance to determine the average size of fluctuations, $\overline{\delta Q}$, for an ensemble of pair instances.  The exact way to do this would be to evaluate weighted average
\begin{equation}
    \overline{\delta Q} = e\sqrt{\alpha(1-\alpha)}\int_{-\infty}^\infty \int_{-\infty}^\infty \sqrt{N_\shortleftarrow+N_\shortrightarrow} P(N_\shortrightarrow)P(N_\shortleftarrow) dN_\shortrightarrow dN_\shortleftarrow.
    \label{SI:ickyIntegral}
\end{equation}
Conveniently, we can use a shortcut to avoid evaluating this integral.  In our simulations, the number of sites is large (and even more so in a real experiment). This means that fluctuations of $N_\rightleftarrows$ (given by $\sigma_\rightleftarrows = \sqrt{Np(1-p)(1-p(1-p))}$) are small compared to the mean (given by $\overline{N_\rightleftarrows}=Np(1-p)$).  Therefore the only region where the contributions to the integral in Suppl.~Eq.~\ref{SI:ickyIntegral} are appreciable is near to the mean; accordingly, we can simply replace $N_\rightleftarrows$ by $\overline{N_\rightleftarrows}$ in Suppl.~Eq.~\ref{SI:deltaQ_1}, which gives Eq.~3 of the main text, \textit{i.e.},
\begin{equation}
\overline{\delta Q} = e\frac{L}{l_0}\sqrt{2p(1-p)\alpha(1-\alpha)}.
\label{SI:eq:first_flucts}
\end{equation}
In Suppl.~Fig.~\ref{SI:fig:FluctSweeps}, we run many simulations to plot $\overline{\delta Q}$ \textit{vs.}~$L/l_0$ for $l=l_0$ and $l=10 l_0$ and several values of $\alpha$ and $p$. Rescaling by the factor, $\sqrt{2p(1-p)\alpha(1-\alpha)}$, we confirm we collapse all of our numerically simulated data (Suppl.~Fig.~\ref{SI:fig:FluctSweeps}e).

\section{Nucleation}
\label{SI:Nucleation}
In Fig.~4 of the main paper, a physically-derived nucleation process is used to generate heterogeneous surfaces, which can then be characterized in terms of donor probability $p$ and length scale $l/l_0$ and used in contact charging simulations.
This process is formulated from general energy considerations, and without assuming the nature of the sites.
The main assumption of this model is that the probability to form a site is dependent on the presence of neighboring sites of the same type.
Indeed, the length scales observed in the experiment seem to indicate that donor and acceptor sites tend to form patches of more than one site.
One possible explanation is that this situation is more energetically favorable than randomly distributed sites.
We will therefore assume that the probability for a site to form increases with the number of neighboring sites of the same type.

\begin{figure}
    \centering
    \includegraphics[scale=.8]{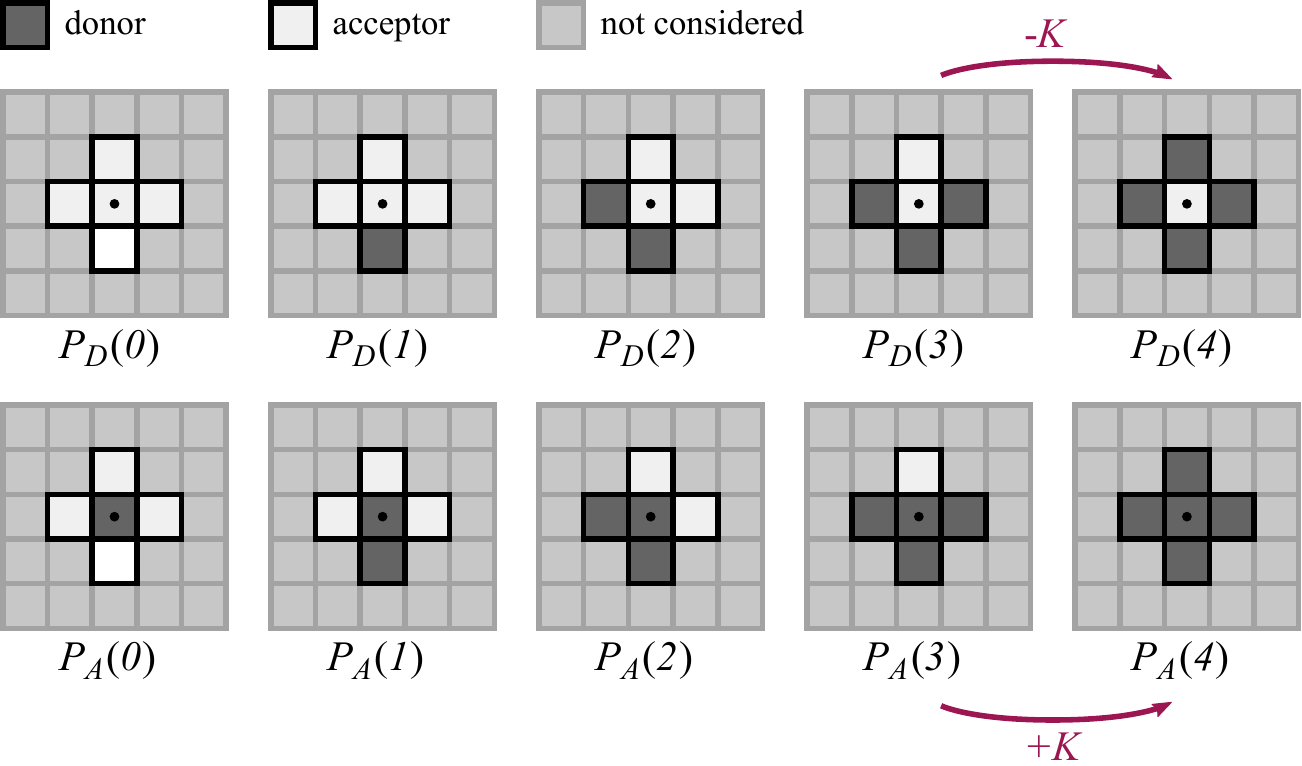}
    \caption{We assume that the probability for a given site (dotted cell) to change state is dependent on the configuration of the first neighbors (Suppl.~Eq.~(\ref{SI:eq:proba})).
    If the site is an acceptor (resp. donor), its probability to become a donor (resp. acceptor) is given by $P_D (\nu)$ (resp. $P_A(\nu)$), where $\nu$ is the number of donors among the first neighbors.
    With one more neighboring donor, the activation energy decreases (resp. increases) by $K$, which in turns increases (resp. decreases) the probability of transition.
    }
    \label{SI:fig:nucl}
\end{figure}

Suppose that the rates at which sites of each type form follows the general Arrhenius-like expression
\begin{equation}
    J = J_0 \exp \left( - E / kT \right)
\end{equation}
where the activation energy $E$ is a function of the number of neighbors.
This rate will determine the probability to form a new site at each time step.
Considering only the first neighbors, we denote $\nu_D = \nu \leq 4$ the number of neighboring donors, $\nu_A = 4-\nu$ the number of neighboring acceptors, and $P_A (\nu)$ ($P_D (\nu)$) the probability for an acceptor (donor) to form on the surface.
For example, we can consider that one species on the surface could act as a donor site, and the absence of that species would define an acceptor site.
In that case, $P_D (0)$ would correspond to the probability for one such site to form on a cell surrounded by donor sites, while $P_A (0)$ would correspond to the probability for a single isolated donor site to disappear.
We assume that each neighbor affects the activation energy linearly, and denote $K$ the change in energy due to the addition or removal of a single neighboring site.
In our example, $K$ can therefore be understood as a (dimensionless) interaction energy between two adjacent donor sites.
If donors are more likely to form near donors, we can expect $P_D$ to increase monotonously with $\nu$.
Conversely, $P_A$ should decrease monotonously with $\nu$.
We therefore propose the general expression
\begin{equation}
\begin{aligned}
    P_A (\nu) &= P_{A,\,0} \exp \left( - K \nu \right)\\
    P_D (\nu) &= P_{D,\,0} \exp \left( - K (4-\nu) \right).
\end{aligned}
\label{SI:eq:proba}
\end{equation}
In the main paper, we also have $P_{A,\,0} = P_{D,\,0} = P_0$, so that the donor/acceptor probabilities are symmetric, reducing the parameters to just $K$ and $P_0$.
All possible configurations are shown in Suppl.~Fig.~\ref{SI:fig:nucl}.

\subsection{Protocol}

The characteristics of a surface are its total size $L/l_0$, the typical size of the patches $l/l_0$ and the probability for a site to be a donor $p$.
First, a matrix of size $L/l_0 \times L/l_0$ is randomly filled with donor and acceptor sites.
The output value of $p$ will be determined by the equilibrium between the creation of new donor and acceptor sites.
Assuming $P_{A,\,0} = P_{D,\,0} = P_0$, we can expect $p_{out} \approx 0.5$.
We therefore choose $p_{in} = 0.5$ in order to reach the dynamic equilibrium faster.
The patch size $l/l_0$ is influenced by $K$, $P_{A,\,0}$ and $P_{D,\,0}$.
While measuring $p$ is straightforward, determining $l/l_0$ is not trivial.
The spatial correlation function
\begin{equation}
    C(r) = \langle s(R) s(R+r) \rangle - \langle s(R) \rangle\langle s(R+r) \rangle
    \label{SI:eq:correl}
\end{equation}
is calculated on the surface, with $s=1$ for a donor and $0$ for an acceptor.
The first zero crossing therefore corresponds to the typical correlation length between sites.

\begin{figure}
    \centering
    \subfigure{\includegraphics[scale=.95]{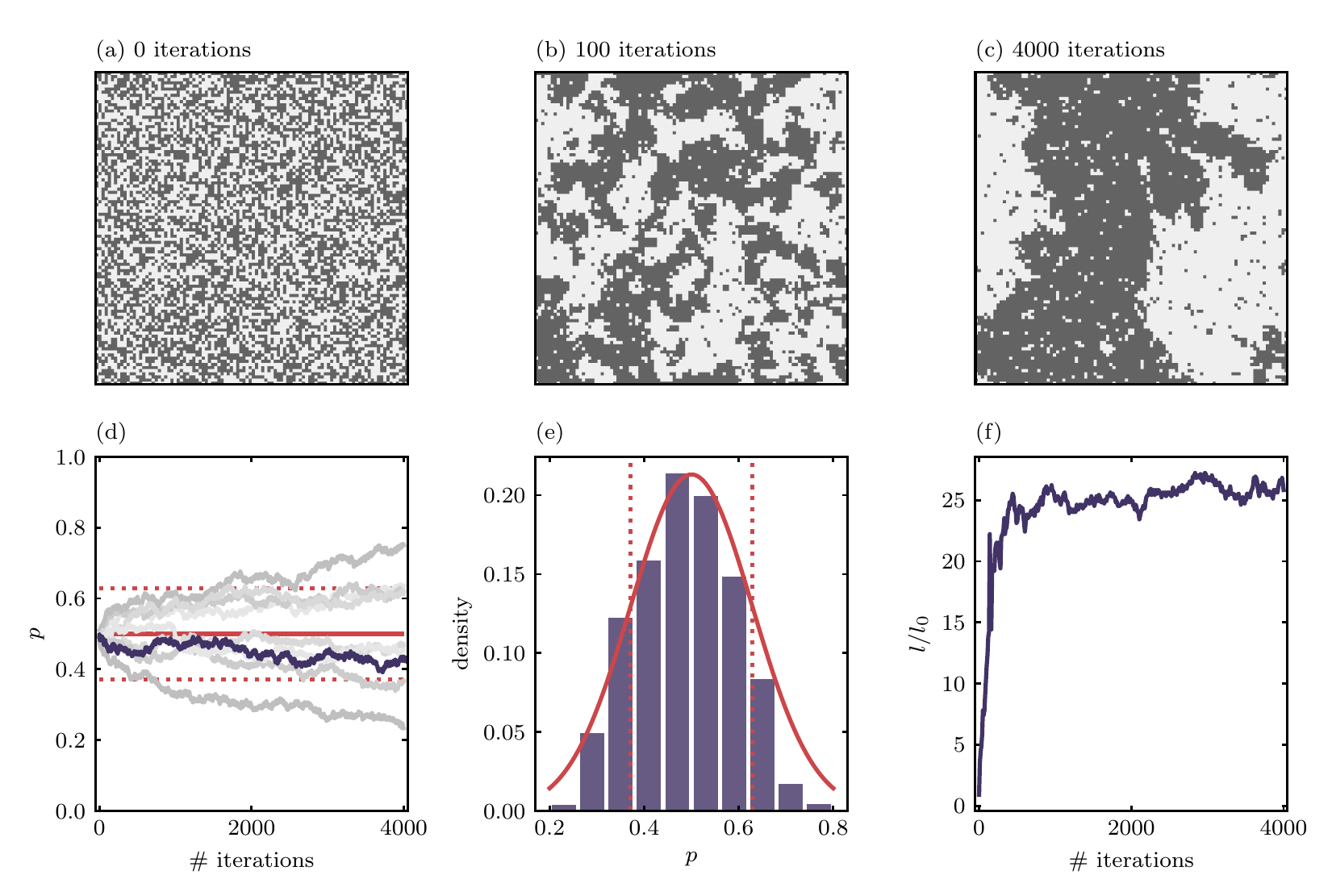}}
    \caption{Successive steps to form a surface through a generic nucleation process. (a) A $100\times 100$ matrix is filled with donors and acceptors, randomly assigned with a probability of 0.5. (b)-(c) After several iterations, with $P_0 = 0.5$ and $K=0.8$, features emerge on the surface. (d) On average, the donor probability $p$ is still 0.5. The typical scale of variation between two surfaces is given by $\sigma_p = l/L \sqrt{p(1-p)}$ (dotted lines). To illustrate this, the evolution of $p$ is shown for 8 other surfaces, in light grey. (e) After enough iterations (here, 350000), a given surface will have explored all of the possible values for $p$. The solid line is a Gaussian centered in $p=0.5$ with standard deviation $\sigma_p$. (f) The size of the patches $l/l_0$, measured through the correlation function from Eq.~\ref{SI:eq:correl}, grows and reaches an equilibrium.}
    \label{SI:fig:protocol}
\end{figure}

A typical run consists in the following steps.
The input parameters are the change in activation energy due to a single neighbor $K$ and the probability $P_0$.
\begin{enumerate}
    \item Either create a random surface of size $L/l_0$ and probability $p_{in}$ or start from a previously generated surface. In the main paper, we compute increasing values of $K$ and use the previous surface as a starting point.
    \item For each cell on the surface: 
    \begin{enumerate}
        \item count the number of neighbors $\nu$;
        \item calculate the probability to change state ($P_A (\nu)$ or $P_D (\nu)$);
        \item either change state or not.
    \end{enumerate}
    \item Repeat step 2 until an equilibrium is reached (5000 iterations in the main paper).
    \item Determine $p$ and calculate the correlation function $C(r)$ as defined in Suppl. Eq.~\ref{SI:eq:correl}. 
    \item Determine $l/l_0$ by averaging $C(r)$ over several surfaces and considering the first zero crossing within the standard deviation (see Fig.~4 of the main paper).
\end{enumerate}
The steps of a given run are shown in Suppl.~Fig.~\ref{SI:fig:protocol}.
The parameters are $L/l_0 = 100$, $p_{in} = 0.5$, $P_0 = 0.5$ and $K=0.8$.
In Suppl.~Fig.~\ref{SI:fig:protocol}a-c, we see the emergence of a length scale on the surface.
This increases the typical fluctuations of $p$, which is given by $\sigma_p = l/L \sqrt{p(1-p)}$.
Supplementary Fig.~\ref{SI:fig:protocol}d shows the evolution of $p$ for a few example surfaces.
On a long enough timescale, a given surface will explore all of the possible values of $p$, which can be seen in Suppl.~Fig.~\ref{SI:fig:protocol}e.
The evolution of the length scale $l/l_0$ is given in Suppl.~Fig.~\ref{SI:fig:protocol}f.
It starts close to unity, and grows to reach an equilibrium value after typically 100 to 1000 iterations.
This equilibrium value of $l/l_0$ depends on the input parameters, particularly $K$.

\section{Charging with multiple length scales}

\begin{figure}
    \centering
    \includegraphics[]{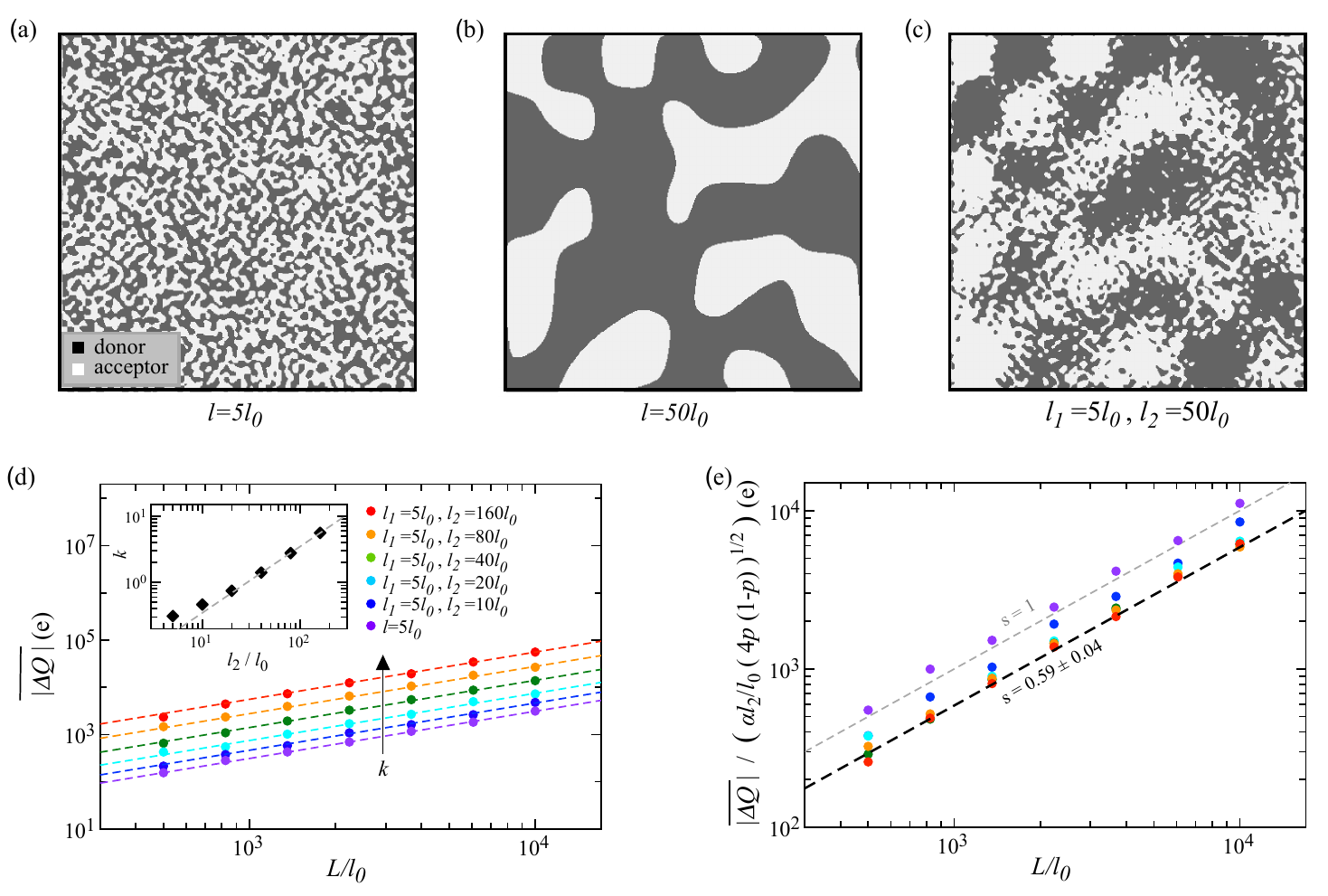}
    \caption{Charging with multiple length scales.  (a) A surface with a single length scale at $l=5l_0$ ($p=0.5$).  (b) A surface with a single length scale at $l=50l_0$ ($p=0.5$).  (c) A surface with two length scales, one at $l_1=5l_0$ and a larger at $l_2 = 50l_0$ ($p=0.5$).  (d) Mean absolute charge transferred, $\overline{|\Delta Q|}$, \textit{vs.}~$L/l_0$, between surfaces with two length scales, a smaller one at
    $l_1=5l_0$ a larger one as indicated in the legend ($p=0.5$ and $\alpha=0.1$ in all cases).  The scale of charge exchange clearly grows with larger scale, $l_2$.  Fitting each curve (dashed colored lines) to $\overline{|\Delta Q|}=kL/l_0$ shows that the prefactor, $k$, is proportional to $l_2/l_0$ for $l_2>>l_1$ (inset, where black dashed line represents $k\propto l_2/l_0$).  (e) Rescaling all data by the factor $\alpha l_2/l_0 \sqrt{4p(1-p)}$ collapses data in the $l_2>>l_1$ regime, albeit with a prefactor, $s$, that is less than 1.  This indicates that, while charge transfer is dominated by larger donor/acceptor length scales if multiple length scales are present, the efficiency is somewhat reduced.  
    }
    \label{SI:fig:doubleL}
\end{figure}

The KPFM data by Baytekin, \textit{et al.}~\cite{SI:Baytekin:2011bx}, reveals two length scales for the patches: one at $\sim$45 nm, and another at $\sim$450 nm.  This brings up a natural question:  if two (or more) length scales are present at the meso scale, how is the charging behavior affected?  Here we will argue that the larger length scales dominate.  We generate a surface with two donor/acceptor length scales by populating a matrix with normal random values on two sub-matrices, one at every ($l_1/l_0$)\ts{th} point, and another at ($l_2/l_0$)\ts{th} point ($l_1<l_2$).  Interpolating between these points at the $l_0$ scale and thresholding as before, we create surfaces such as shown in Suppl.~Fig.~\ref{SI:fig:doubleL}c.  This is one of many ways to introduce two length scales---looking into all such possible ways is beyond the scope of our work.  Qualitatively, however, this procedure (1) preserves the feature $p_{meas} = p_{in}$ and (2) creates surfaces where the smaller donor (acceptor) patches `intrude' into the larger acceptor (donor) patches.  In this latter point, it is qualitatively similar to the KPFM data presented by Ref.~\cite{SI:Baytekin:2011bx}.

In Suppl.~Fig.~\ref{SI:fig:doubleL}d, we show how $\overline{|\Delta Q|}$ is affected when the contacting surfaces have two length scales.  Holding the smaller scale, $l_1$, fixed and increasing the larger one, we see that the magnitude of charging increases for increasing $l_2$.  When $l_2>>l_1$, the growth in $\overline{|\Delta Q|}$ is approximately linear with $l_2/l_0$ (inset to Suppl.~Fig.~\ref{SI:fig:doubleL}d), which indicates that if the scales are sufficiently separated then the larger one dominates.  In Suppl.~Fig.~\ref{SI:fig:doubleL}e, we attempt to collapse our data by the previously found factor $\alpha l_2/l_0 \sqrt{4p(1-p)}$ (but where we've replaced $l$ by $l_2$).  This collapses all data for $l_2>>l_1$, but with a prefactor that is lower than in the single length scale case.  Qualitatively, this can be explained by considering that the larger donor (acceptor) patches are effectively smaller than $l_2$ given the intrusive acceptor (donor) regions of size $\sim l_1$ that now exist inside of them.  Nonetheless, under the conditions we examine here this effect is relatively small, and to within a factor of order 1 our model can still account for the charging behavior simply by assuming one length scale at $l=l_2$.

\begin{figure}
    \centering
    \includegraphics[scale=.95]{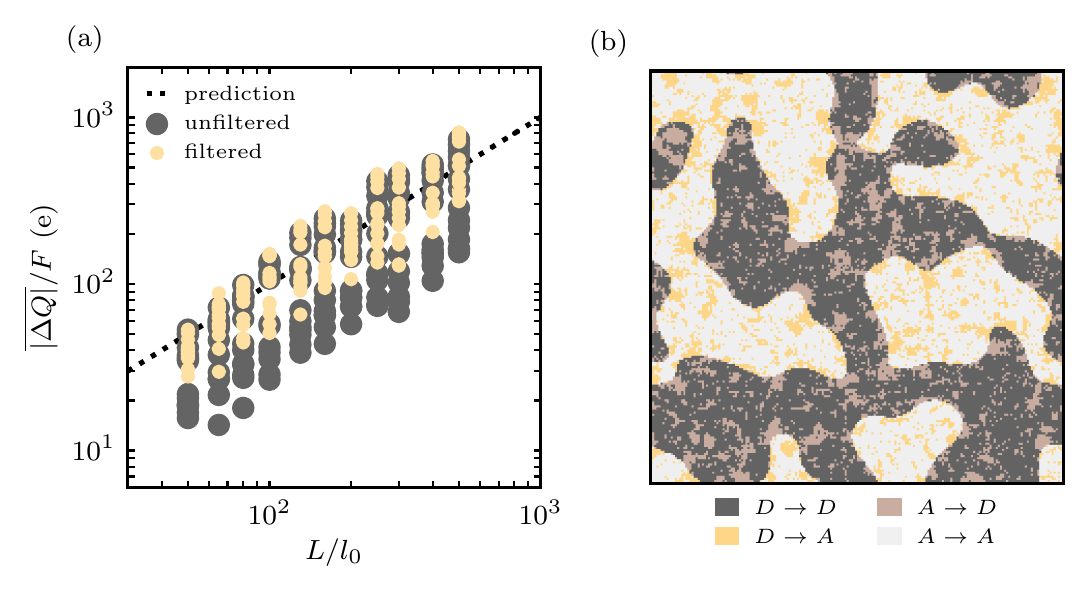}
    \caption{(a) The collapse of charging data from Fig.~4 in the main text is improved when a filter is applied to the surfaces, eliminating features smaller than the measured length scale. The filter is a low-pass Butterworth filter of order 5 with a cutoff in frequency domain that corresponds to $l/l_0$. (b) Example of a surface after filtering. $A\rightarrow A$ and $D\rightarrow D$ represent respectively donor and acceptor sites left unchanged by the filter, while $A\rightarrow D$ and $D\rightarrow A$ represent smaller features that were removed by the filter.}
    \label{SI:fig:filter}
\end{figure}

The presence of multiple length scales might explain the imperfect collapse from Fig.~4e in the main text. For some values of $l/l_0$, the measured $\overline{|\Delta Q|}$ is smaller than the prediction by a factor of order 1, similarly to what we see in Suppl.~Fig.~\ref{SI:fig:doubleL}e. To determine if it is due to the presence of multiple length scales, we remove features smaller than the dominant length scale on each surface, then perform the same single-contact measurement again. To remove the small features, we multiply the Fourier transform of every surface by a Butterworth low-pass filter of order 5 where the cutoff is set at the wavenumber that corresponds to the dominant length scale. In Suppl.~Fig.~\ref{SI:fig:filter}a, we compare the collapse of $\overline{|\Delta Q|}$ by the prefactor $F=\alpha l/l_0 \sqrt{4p(1-p)/\pi}$ with and without the filter. One can see that the collapse is significantly improved by the removal of features smaller than $l/l_0$, which suggests the presence of smaller length scales that are not accounted for in the model. An example of filtering is shown in Suppl.~Fig.~\ref{SI:fig:filter}a, where the filtered surface (shades of gray) is superimposed with the original surface. The sites that have been affected by the filter are shown in light brown (acceptor to donor) and yellow (donor to acceptor).

\section{Reconsideration of previous experimental results}

\begin{figure}
    \centering
    \includegraphics[scale=.95]{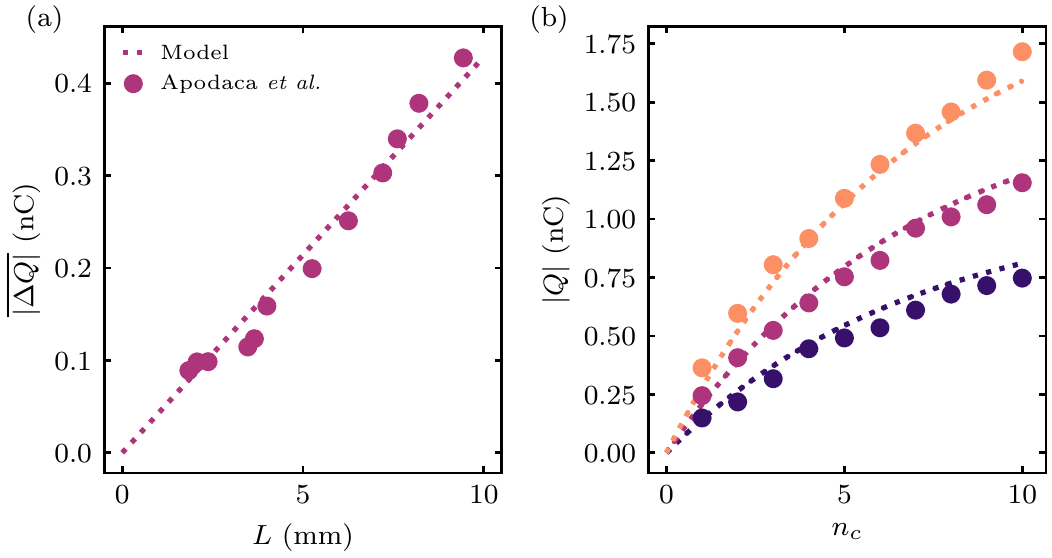}
    \caption{Comparison with the experiments from Apodaca \emph{et al.}~\cite{SI:Apodaca:2009dr}. (a) We can find the length scale $l_0$ through a fit on the single-contact data. (b) The average charge after multiple contacts follows an exponential saturation, with each curve corresponding to a surface with a different area. We use the same coefficient $\alpha=0.143$ for each curve.}
    \label{SI:fig:exp}
\end{figure}

In this section, we compare our model with the experimental data from Apodaca \emph{et al.}~\cite{SI:Apodaca:2009dr}. Two types of experiments have been performed: single-contact, and multiple successive contacts, as show in Suppl. Fig.~\ref{SI:fig:exp}a and~\ref{SI:fig:exp}b, respectively. In multiple-contact experiments, the charge increases progressively until reaching a plateau, as described by Suppl. Eq.~\ref{SI:eq:multi_predict} and~\ref{SI:eq:continuous3}. In Supplementary Fig.~\ref{SI:fig:exp}b, we fit the continuous expression from Suppl. Eq.~\ref{SI:eq:continuous3} to the experimental data from~\cite{SI:Apodaca:2009dr}. We find $\alpha = 0.143$ and $Q_f = [ 0.972, 1.489, 2.187]$, where each value for $Q_f$ correspond to a different area for the surface (respectively 16.00, 35.22 and 61.62~mm$^2$). We verify that $Q_f \propto \sqrt{A}$. The value of $\alpha$ obtained can then be used to determine the length scale $l_0$ using the single-contact data in Suppl. Fig.~\ref{SI:fig:exp}a. Indeed, the average charge after a single contact is given by Suppl. Eq.~\ref{SI:eq:l_eq_l} and depends on $L$, $l_0$, $l$, $p$ and $\alpha$. We already know $L$ and $\alpha$, and the KPFM data from Baytekin \emph{et al.}~\cite{SI:Baytekin:2011bx} gives us a value for $l$. Baytekin \emph{et al.} identified two length scales, 450~nm and 44~nm. However, as established in the previous section, charging is dominated by the larger length scale, granted that they are sufficiently separated. We therefore consider here that $l=450$~nm. Finally, we make the assumption that donors and acceptors are equally probable, \emph{i.e.} $p=0.5$. The fit of Eq.~\ref{SI:eq:l_eq_l} on the data from~\cite{SI:Apodaca:2009dr} shown in Suppl. Fig.~\ref{SI:fig:exp}a gives us $l_0 = 3.67$~\r{A}. In other words, a square of side $\sim 4$~\r{A} would donate/accept a single elementary charge. By comparison, if we do not consider the influence of donor/acceptor patches, we can fit the data using Eq.~\ref{SI:eq:l_eq_l0} which yields $l_0 = 0.003$~\r{A}. Similarly, the value found in~\cite{SI:Apodaca:2009dr} corresponds to $l_0 = 0.005$~\r{A}, differing slightly due to the different $\alpha$ dependency. This would imply an unreasonably high number of charges ($\sim 10^5$) on an area the size of an atom. Furthermore, one can verify that no combination of $\alpha$ and $p$ would yield a plausible value for $l_0$, indicating that the introduction of a length scale $l>l_0$ is necessary.

\newpage
\section{Description of the Supplemental Movie}

A movie is included which depicts the evolution of a surface generated using a physically-derived nucleation process (as described in the Nucleation section of the Supplemental Material). A grid is randomly filled with donors and acceptors ($p=0.5$), then we let the system evolve for 1000 iterations. Parameters are $K=1$, $P_0 = 0.5$ and $L/l_0 = 200$. Features appear on the grid which grow over time until a dynamic equilibrium is reached.